\documentclass[fleqn,usenatbib]{mnras}

\usepackage[T1]{fontenc}
\usepackage{newtxtext,newtxmath}

\DeclareRobustCommand{\VAN}[3]{#2}
\let\VANthebibliography\thebibliography
\def\thebibliography{\DeclareRobustCommand{\VAN}[3]{##3}\VANthebibliography}

\usepackage{graphicx}	
\usepackage{bm}
\usepackage{amsmath}	
\usepackage{multicol, blindtext} 

\usepackage{booktabs}
\usepackage{textcomp}
\usepackage{caption}
\usepackage{xcolor}
\usepackage{soul}
\usepackage{subcaption}

\defcitealias{trindadefalcao2021a}{Paper 1}

\title[\textit{Hubble Space Telescope} Observations of Nearby Type 1 Quasars]{\textit{Hubble Space Telescope} Observations of Nearby Type 1 Quasars. I. Characterisation of the Extended {[O~III]} 5007\AA~Emission}

\author[Trindade Falcão et al.]{
Anna Trindade Falcão ,$^{1,2}$\thanks{E-mail: anna.trindade\_falcao@cfa.harvard.edu}
S. B. Kraemer,$^{2}$
T. C. Fischer,$^{3}$
H. R. Schmitt,$^{4}$
L. Feuillet,$^{2}$
D. M. Crenshaw,$^{5}$
\newauthor
M. Revalski,$^{6}$
W. P. Maksym,$^{7}$
M. Vestergaard,$^{8,15}$
M. Elvis,$^{1}$
C. M. Gaskell,$^{9}$
L. C. Ho,$^{10}$
H. Netzer,$^{11}$
\newauthor
T. Storchi-Bergmann,$^{12}$
T. J. Turner,$^{13}$
M. J. Ward$^{14}$
\\
$^{1}$Harvard-Smithsonian Center for Astrophysics, 60 Garden St., Cambridge, MA 02138, USA\\
$^{2}$Institute for Astrophysics and Computational Sciences and Department of Physics, The Catholic University of America, Washington, DC 20064, USA\\
$^{3}$AURA for ESA, Space Telescope Science Institute, 3700 San Martin Drive, Baltimore, MD 21218, USA\\
$^{4}$Naval Research Laboratory, Washington, DC 20375, USA\\
$^{5}$Department of Physics and Astronomy, Georgia State University, Astronomy Offices, 25 Park Place, Suite 600, Atlanta, GA 30303, USA\\
$^{6}$Space Telescope Science Institute, 3700 San Martin Drive, Baltimore, MD 21218, USA\\
$^{7}$NASA Marshall Space Flight Center, Huntsville, AL 35812, USA\\
$^{8}$DARK, Niels Bohr Institute, University of Copenhagen, Jagtvej 155A, 2200 Copenhagen N, Denmark\\
$^{9}$Department of Astronomy and Astrophysics, University of California, Santa Cruz, CA 95064, USA\\
$^{10}$Kavli Institute for Astronomy and Astrophysics, Peking University; School of Physics, Department of Astronomy, Peking University; Beijing 100871, China\\
$^{11}$School of Physics and Astronomy, Tel Aviv University, Tel Aviv 69978, Israel\\
$^{12}$Departamento de Astronomia, Universidade Federal do Rio Grande do Sul, IF, CP 15051, 91501-970 Porto Alegre, RS, Brazil\\
$^{13}$Eureka Scientific, Inc., 2452 Delmer Street, Suite 100, Oakland, CA 94602-3017, USA\\
$^{14}$Centre for Extragalactic Astronomy, Department of Physics, University of Durham, South Road, Durham DH1 3LE, UK\\
$^{15}$Department of Astronomy and Steward Observatory, 933 Cherry Avenue, Tucson, AZ 85721, USA
}

\date{Accepted XXX. Received YYY; in original form ZZZ}

\pubyear{2023}
\begin{document}
\label{firstpage}
\pagerange{\pageref{firstpage}--\pageref{lastpage}}
\maketitle

\begin{abstract}
We use the \textit{Hubble Space Telescope} to analyse the extended [O~III] 5007\AA~emission in seven bright radio-quiet type 1 quasars (QSO1s), focusing on the morphology and physical conditions of their extended Narrow-Line Regions (NLRs). We find NLRs extending 3-9 kpc, with four quasars showing roughly symmetrical structures ($b/a$=1.2-1.5) and three displaying asymmetric NLRs ($b/a$=2.4-5.6). When included with type 1 and type 2 AGNs from previous studies, the sizes of the extended [O~III] regions scale with luminosity as $R_{\rm [OIII]}\sim L_{\rm [OIII]}^{0.5}$, consistent with photoionisation. However, when analysed separately, type 1s exhibit a steeper slope ($\gamma_{1}$=0.57$\pm$0.05) compared to type 2 AGNs ($\gamma_{2}$=0.48$\pm$0.02). We use photoionisation modeling to estimate the maximum NLRs sizes, assuming a minimum ionisation parameter of log$(U)=-3$, an ionising luminosity based on the $L_{\rm [O~III]}$-derived bolometric luminosity, and a minimum gas number density $n_{\rm H}\sim100$~cm$^{-3}$, assuming that molecular clouds provide a reservoir for the ionised gas. The derived sizes agree well with direct measurements for a sample of type 2 quasars, but are underestimated for the current sample of QSO1s. A better agreement is obtained for the QSO1s using bolometric luminosities derived from the 5100\AA~continuum luminosity. Radial mass profiles for the QSO1s show significant extended mass in all cases, but with less [O~III]-emitting gas near the central AGN compared to QSO2s. This may suggest that the QSO1s are in a later evolutionary stage than QSO2s, further past the blow-out stage.
\end{abstract}

\begin{keywords}
galaxies: active -- quasars: emission lines
\end{keywords}


\section{Introduction}
\label{sec:intro}

The evolution and growth of massive galaxies are intricately connected to their Active Galactic Nuclei (AGN) through the interplay between gas feeding \citep[e.g.,][]{storchi2019a} into the nuclear super massive black hole (SMBH), star-formation activity, and AGN feedback to the host galaxy. These are key to explaining \citep[e.g.,][]{fabian2012a, king2015a} present-day empirical scaling relations between the mass of the SMBH and properties of the host \citep[e.g.,][]{merritt2001a,gultekin2009a}, the halting of star formation and death of massive galaxies \citep[e.g.,][]{behroozi2019a}.

There are three main forms for AGN feedback onto the host interstellar medium (ISM): radiation, jets and winds. There is plenty of radiation available, but most of it escapes. Jets are powerful, relativistic and highly collimated, but only occur in up to $\sim$10\% of the AGN population \citep{rafter2009a}. Meanwhile, winds are present in many, and perhaps most, AGN \citep[e.g.,][]{mullaney2013a, woo2016a}, and are potentially universal in radio-quiet sources. AGN winds may be observed as blueshifted UV and X-ray absorption lines, moving with relativistic velocities, up to $\sim$0.3$c$, within tens of parsecs from the central SMBH \citep{crenshaw2003a, tombesi2010a, kraemer2012a, king2015a}. These may also be observed as ionised emission-line gas in the narrow-line region (NLR), on larger scales (100s–1,000s of pc), and with velocities reaching up to $\sim$2,000 km~s$^{-1}$ \citep{crenshaw2005a, crenshaw2010a, fischer2013a, bae2016a, revalski2018a, trindadefalcao2021a,molina2022a}.

\citet{fischer2018a} characterized the [O~III] 5007\AA~gas in the NLR of twelve nearby and obscured quasars. They used \textit{HST}/STIS to examine the extent and kinematics of the ionised gas, reporting the presence of ionised outflows in all twelve targets of their sample, with an average maximum outflow radius of $\sim$600 pc. Our follow-up study on the same sample (\citealt{trindadefalcao2021a}, hereafter \citetalias{trindadefalcao2021a}) was focused on the characterisation of the physical properties and energetics of these outflows. Our results showed that the ionised outflowing winds in the sample lack the necessary kinetic power to deliver effective feedback to their host galaxy \citep{dimatteo2005a, hopkins2010a}. These results, combined with the findings of other similar high-resolution kinematics studies of nearby Seyferts \citep[e.g.,][]{das2005a, crenshaw2010a, storchi2010a, fischer2018a, revalski2021a}, question the effectiveness of AGNs in being capable of clearing ISM material from their hosts.

In the current work, we present \textit{HST}/ACS imaging of seven of the brightest (log~$L_{\rm bol}>46$~erg~s$^{-1}$) PG quasars from the Bright QSO Survey, obtained through Hubble Program ID 15281 (PI: Kraemer), at $\sim$200 pc resolution. We map the morphology of the extended [O~III] emission as a function of distance from the central SMBH, and examine the relation between the extent and luminosity of their [O~III] NLRs. By extending our analysis to such sample of nearby, unobscured quasars, we aim to shed light on the presence and nature of AGN feedback processes in quasars, potentially revealing how these objects evolve over time.

The current paper is organized as follows: Section \ref{sec:sample, observations} discusses our observations, the sample, data reduction, and analysis; Section \ref{sec:results} presents our results on the morphology of the ionised NLRs in individual targets; Section \ref{sec:discussion} discusses these results, including the relation between ionised NLR sizes and luminosities (Section \ref{sec:relation_r_and_L}),  our models for estimating ionised NLR extents (Section \ref{sec:Constraints_on_NLR_Sizes}), bolometric luminosity calculation (Section \ref{sec:bolometric_luminosity}), and ionised gas mass profiles (Section \ref{sec:Ionised Gas Masses}); finally, Section \ref{sec:conclusions} presents our conclusions. 

For this sample, we have also obtained \textit{HST}/STIS spectroscopic observations of the ionised gas. All the spectral and kinematic analysis will be presented in a subsequent paper.

Throughout this work, we adopt \textit{Wilkinson Microwave Anisotropy Probe (WMAP)} 12-year results cosmology, a flat Lambda cold dark matter cosmology with H$_{0}$=71~km~s$^{-1}$~Mpc$^{-1}$, $\Omega_{0}$=0.28, and $\Omega_{\lambda}$=0.72.

\section{Sample, Observations and Analysis}
\label{sec:sample, observations}

\subsection{Sample Selection}
\label{sec:sample_selection}

The PG QSO1s in our sample are listed in Table \ref{tab:targets}, and comprise seven of the brightest (log$L_{\rm bol}>46$~erg~s$^{-1}$) radio-quiet PG quasars $z<0.5$ from the \citet{boroson1992a} sample. Table \ref{tab:targets} also lists their respective positions, redshifts, angular scales, and luminosity distances, obtained from the NASA/IPAC Extragalactic Database (NED). The angular scale and luminosity distances were corrected to the cosmic microwave background radiation reference frame, and the shown uncertainties correspond to the maximum errors arising from the uncertainties in H$_{0}$ and \textit{z}.

\begin{table*}
\begin{center}  
\begin{tabular}{cccccccccc}
&\multicolumn{9}{c}{}\\
\hline
\multicolumn{1}{l}{QSO1s}
&\multicolumn{1}{c}{R.A.}
&\multicolumn{1}{c}{Decl.}
&\multicolumn{1}{c}{$z$}
&\multicolumn{1}{c}{Scale}
&\multicolumn{1}{c}{Distance}
&\multicolumn{1}{c}{Observation}
&\multicolumn{1}{c}{Filter}
&\multicolumn{1}{c}{Exp Time}
&\multicolumn{1}{c}{Band}
\\
 &(2000)&(2000)& &(kpc/$''$)& (Mpc) &Date & & (s) & (rest) \\
\hline
\hline
  PG0026+129 & 00:29:13.702& +13:16:03.89 & 0.140 & 2.39 & 637$\pm$44 &2018-08-30 & FR551N & 1710 & [O~III] \\
  & & & & & & & FR647M & 200 & Continuum 5675\AA \\[.3cm]

  PG0052+251 & 00:54:52.131 & +25:25:38.94 & 0.154 & 2.56 & 678$\pm$48 & 2018-11-06 & FR601N & 1156 & [O~III] \\
  & & & & & & & FR647M & 180 & Continuum 5607\AA \\[.3cm]

  PG0157+001 & 01:59:50.254 & +00:23:41.00 & 0.163 & 2.69  & 717$\pm$50 & 2018-06-12 & FR601N & 1124 & [O~III]\\
  & & & & & & & FR647M & 180 & Continuum 5563\AA\\[.3cm]

  PG0953+414 & 09:56:52.392 & +41:15:22.17 & 0.234 & 3.60  & 1039$\pm$72 & 2018-11-12 & FR601N & 1234 & [O~III] \\
  & & & & & & &FR647M & 180 & Continuum 5243\AA \\[.3cm]

  PG1012+008 & 10:14:54.902 & +00:33:37.55 & 0.187 & 3.03  & 831$\pm$58 & 2018-12-03 & FR601N & 1122 & [O~III]\\
  & & & & & & &FR647M & 180 & Continuum 5451\AA\\[.3cm]

  PG1049–005 & 10:51:51.440 & -00:51:17.73 & 0.360 & 4.87 & 1593$\pm$111 & 2018-11-26 & FR656N & 1684 & [O~III] \\
  & & & & & & &FR647M & 220 & Continuum 4757\AA \\[.3cm]

  PG1307+085 & 09:56:52.392 & +41:15:22.17 & 0.154 & 2.60 & 685$\pm$47 & 2018-12-21 & FR601N & 1132 & [O~III] \\
   & & & & & & &FR647M & 180 & Continuum 5607\AA \\

\hline
\end{tabular}
\end{center}
\caption{\textbf{PG QSO1s Sample: Coverage and Observations Summary.} Note: (1) Galaxy name; (2-3) Galaxy coordinates, from NED; (4) redshift, from NED; (5) scale, from NED; (6) luminosity distance, from NED; (7) date of observation; (8) \textit{HST} filter; (9) exposure time of observation; (10) spectral band.}
\label{tab:targets}
\end{table*}

The choice of redshift threshold ensures that we will map the [O~III] structures on scales of several 100s pc for the most distant targets, while analysing the most luminous quasars ensures that the AGNs in our sample are more luminous than Seyfert galaxies studied in previous works \citep[e.g.,][]{fischer2013a}. The QSO1s in our sample encompass a somewhat wider range in redshift compared to the QSO2s studied in \citetalias{trindadefalcao2021a}, but fall within a similar range in bolometric luminosities (see Section \ref{sec:bolometric_luminosity}).

\subsection{Observations and Data Reduction}
\label{sec:data_reduction}

The observations used in this study were obtained with the Advanced Camera for Surveys (ACS)\footnote{https://www.stsci.edu/hst/instrumentation/acs}/Wide-Field Channel (WFC) ($\sim$0.05$''$/pixel spatial resolution) on board of the \textit{Hubble Space Telescope (HST)}. We employ the FR551N, FR601N, and FR656N filters (Table \ref{tab:targets}) to characterise the morphology of the emission-line gas, and determine the ionised gas mass. For the continuum observations, we use the FR647M filter, selected to encompass a broad continuum region without prominent emission lines (Table \ref{tab:targets}). Data reduction and analysis of both on-band and continuum images follow the standard procedures in the \textit{HST} pipeline. 

We use \texttt{IRAF} (Image Reduction and Analysis Facility) for alignment of on-band and continuum images (\textit{imalign}), and subsequent continuum subtraction (\textit{imarith}). We then divide the images by their respective exposure times, available in the headers, and convert them to units of erg cm$^{-2}$ s$^{-1}$ \AA$^{-1}$ using the header keyword \textit{photflam}. To determine [O~III] 5007\AA~fluxes, we integrate the [O~III] continuum-subtracted images within apertures encompassing the full extent of the optical NLRs in these targets. 

The continuum images are also employed to determine the AGN flux at 5100\AA~(see Section \ref{sec:bolometric_luminosity}). These QSO1s were observed around a rest wavelength of 5500\AA~(Table \ref{tab:targets}), free from significant emission lines, making them particularly suitable for our analysis. For each source, we measure the 5500\AA~continuum fluxes within circular apertures centered at the nuclei, with radii increasing from 1 to 10 pixels, in steps of 1 pixel. These fluxes are used to calculate the fluxes inside annuli of increasing radii with a width of 1 pixel, to determine the light profile of the nuclear emission. We find that, with the exception of PG0157+001, these profiles deviate by less than $<$5\% relative to that of the corresponding filter PSF. After subtracting the emission from the host galaxy, estimated using the emission from the largest annuli, we use our flux measurements and the ACS encircled energy observed through filter F625W, which corresponds to a wavelength similar to that used in our observations\footnote{https://www.stsci.edu/hst/instrumentation/acs/data-analysis/aperture-corrections}, to apply an aperture correction and extrapolate the observed flux at a radius of 5 pixels, to the total flux encircled by a PSF corresponding to the nuclear emission at 5500\AA. PG0157+001 shows a $\sim$30\% excess in the encircled flux value from small to large apertures, consistent with the presence of a significant host galaxy. In this case, we use the measurements from the smaller apertures to determine the AGN flux. The measured fluxes are calibrated using the \texttt{PHOTFLAM} keyword available in the image headers and subsequently converted to the 5100\AA, assuming a spectral slope of $\nu^{-0.5}$ \citep[][]{koratkar1999a}, the wavelength needed for our calculations (see Table \ref{tab:qso1s_lbol_cont}).


\section{Results}
\label{sec:results}

\subsection{Morphology of the Extended Ionised Emission}
\label{sec:morphology}

\begin{figure*}
     \centering
     \begin{subfigure}{0.49\textwidth}
         \centering
        \includegraphics[width=\linewidth]{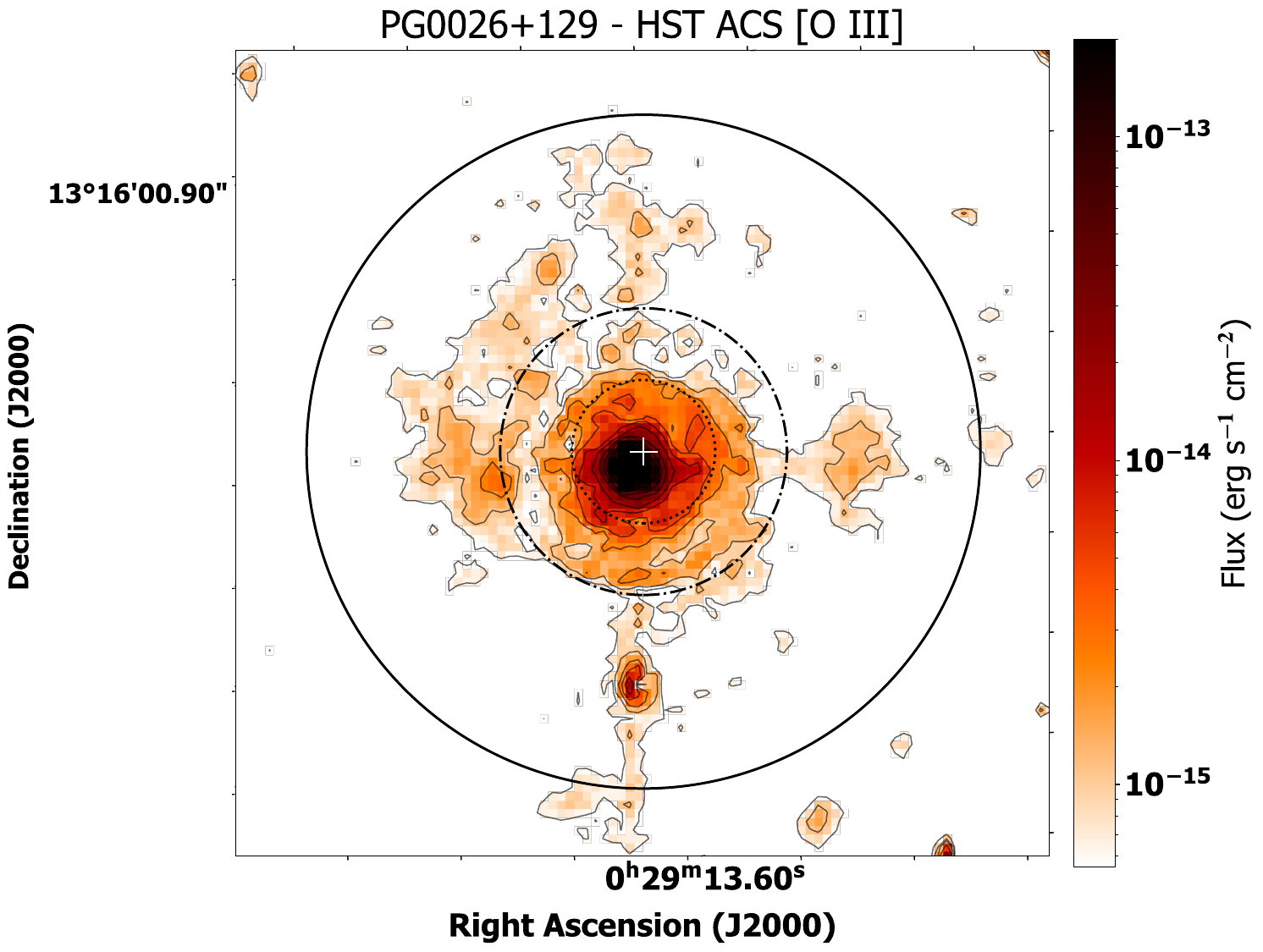}
     \end{subfigure}
     \begin{subfigure}{0.49\textwidth}
         \centering         \includegraphics[width=\linewidth]{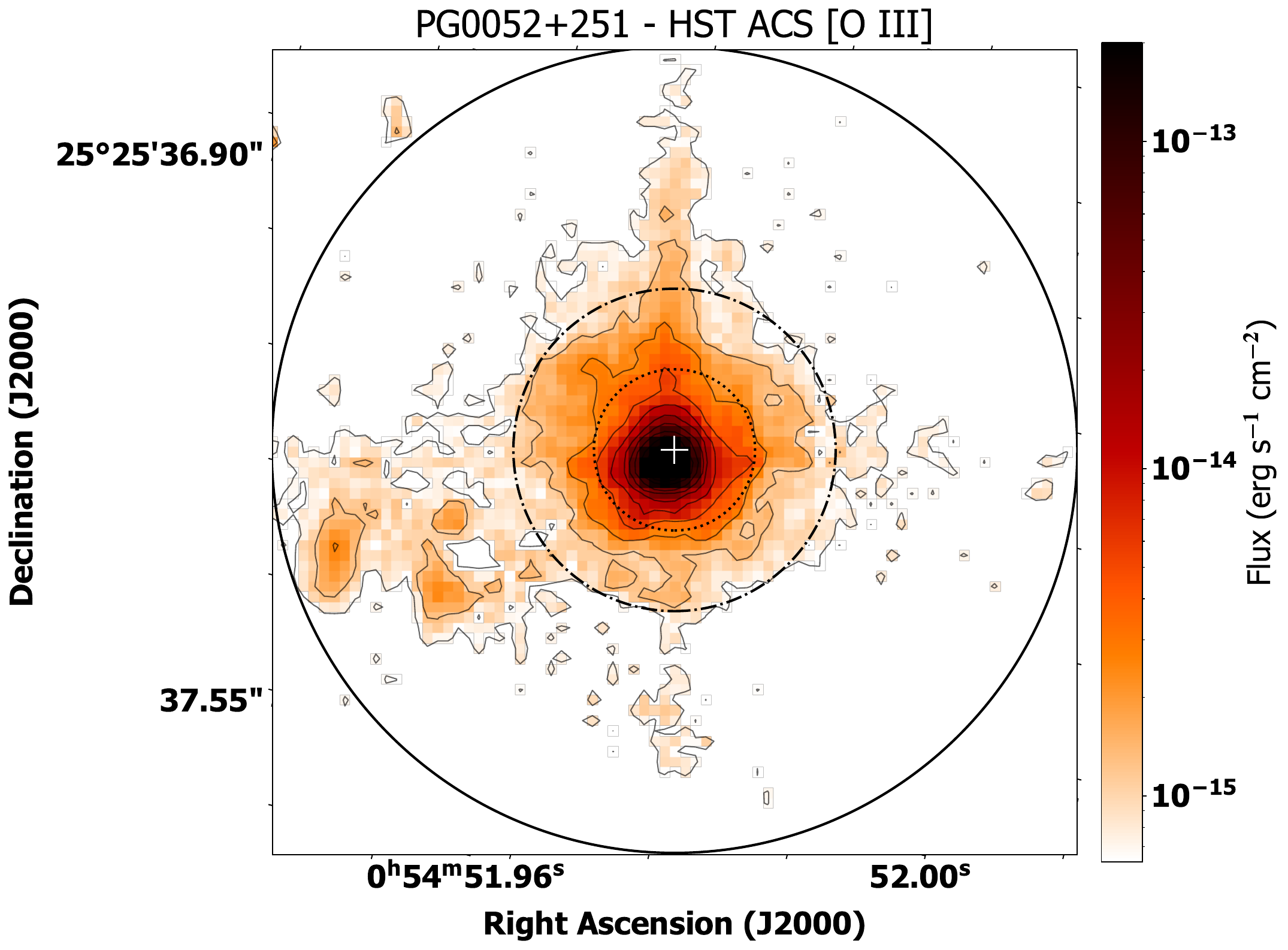}
     \end{subfigure}

     \centering
     \begin{subfigure}{0.49\textwidth}
         \centering
        \includegraphics[width=\linewidth]{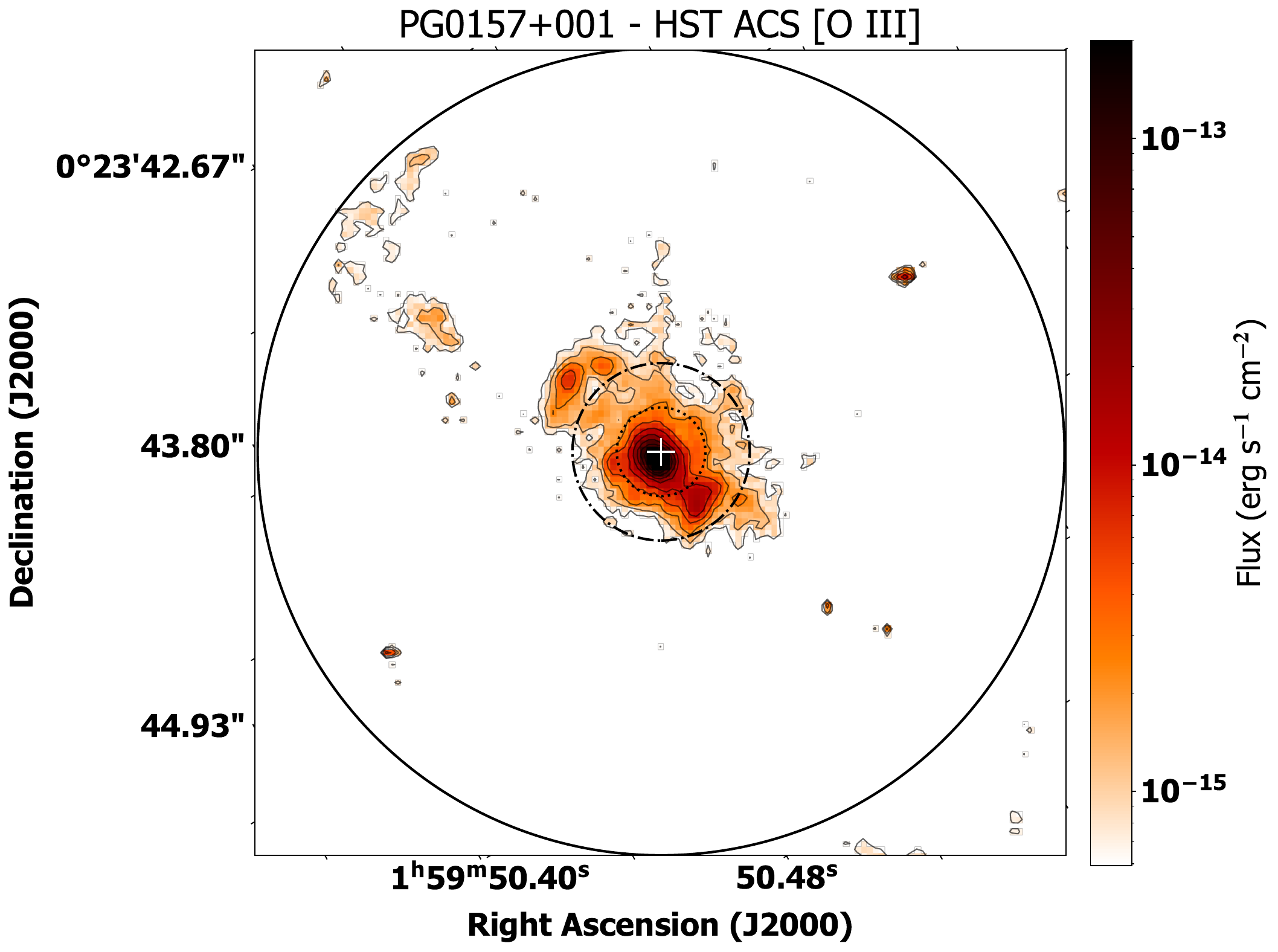}
     \end{subfigure}
     \begin{subfigure}{0.49\textwidth}
         \centering         \includegraphics[width=\linewidth]{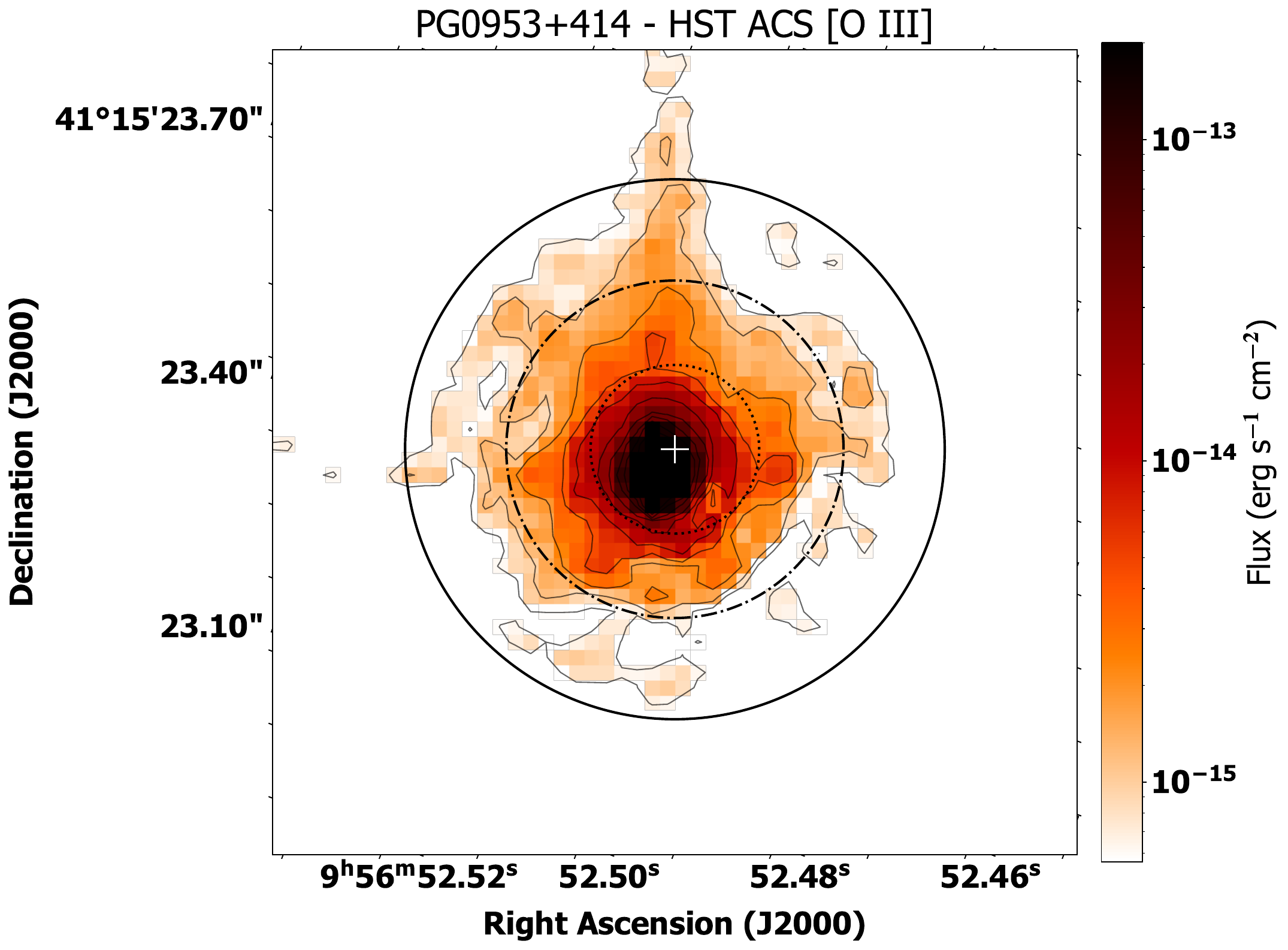}
     \end{subfigure}

     \caption{Continuum-subtracted [O~III] images for the PG QSO1s in the sample. The flux contours start at 3$\sigma_{\rm sky}$ over background, increasing in powers of 3$\sigma_{\rm sky}\times$2$^{n}$, for $n$ up to 10. The nucleus position is measured at the continuum flux peak, and shown as a white cross. Solid lines represent $R_{\rm [O~III]}$, while dashed-dot and dotted lines mark $r=1$~kpc and $r=2$~kpc, respectively. Note that several of the images have diffraction spikes that affect the shapes of the contours.}
     \label{fig:morphology1}
\end{figure*}

The continuum-subtracted [O~III] images for the type 1 quasars in our sample are shown in Fig. \ref{fig:morphology1}. White crosses mark the position of the nuclei, adopted as corresponding to the peak of the continuum flux in each image. The fluxes shown in each panel range from 3$\sigma_{\rm sky}$ to the the maximum flux value in the galaxy. We overplot $n=$10 equally spaced contours, ranging from 3$\sigma_{\rm sky}$ to 3$\sigma_{\rm sky}\times$2$^{n}$. 

\addtocounter{figure}{-1}
\begin{figure*}
     \centering
     \begin{subfigure}{0.49\textwidth}
         \centering
        \includegraphics[width=\linewidth]{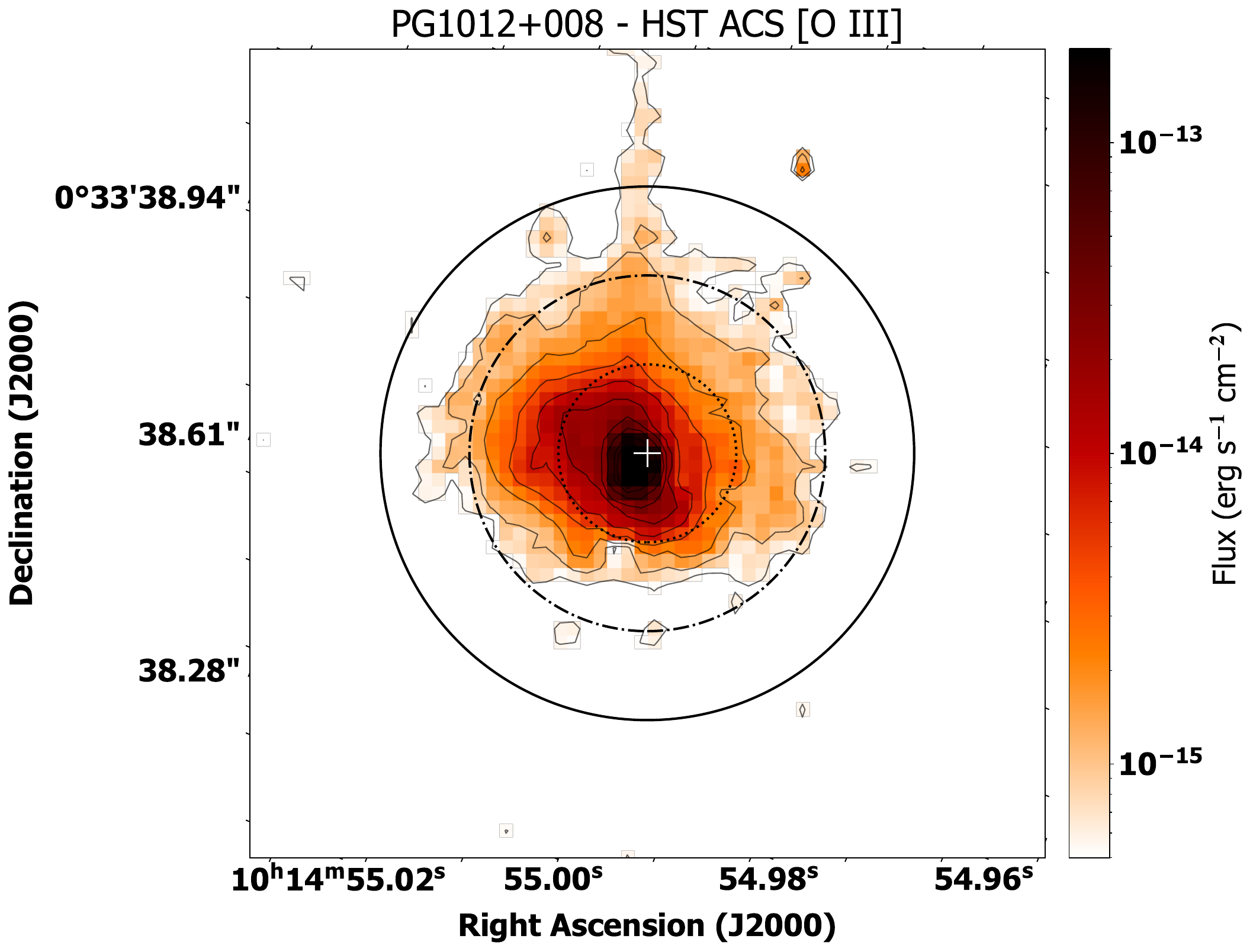}
     \end{subfigure}
     \begin{subfigure}{0.49\textwidth}
         \centering         \includegraphics[width=\linewidth]{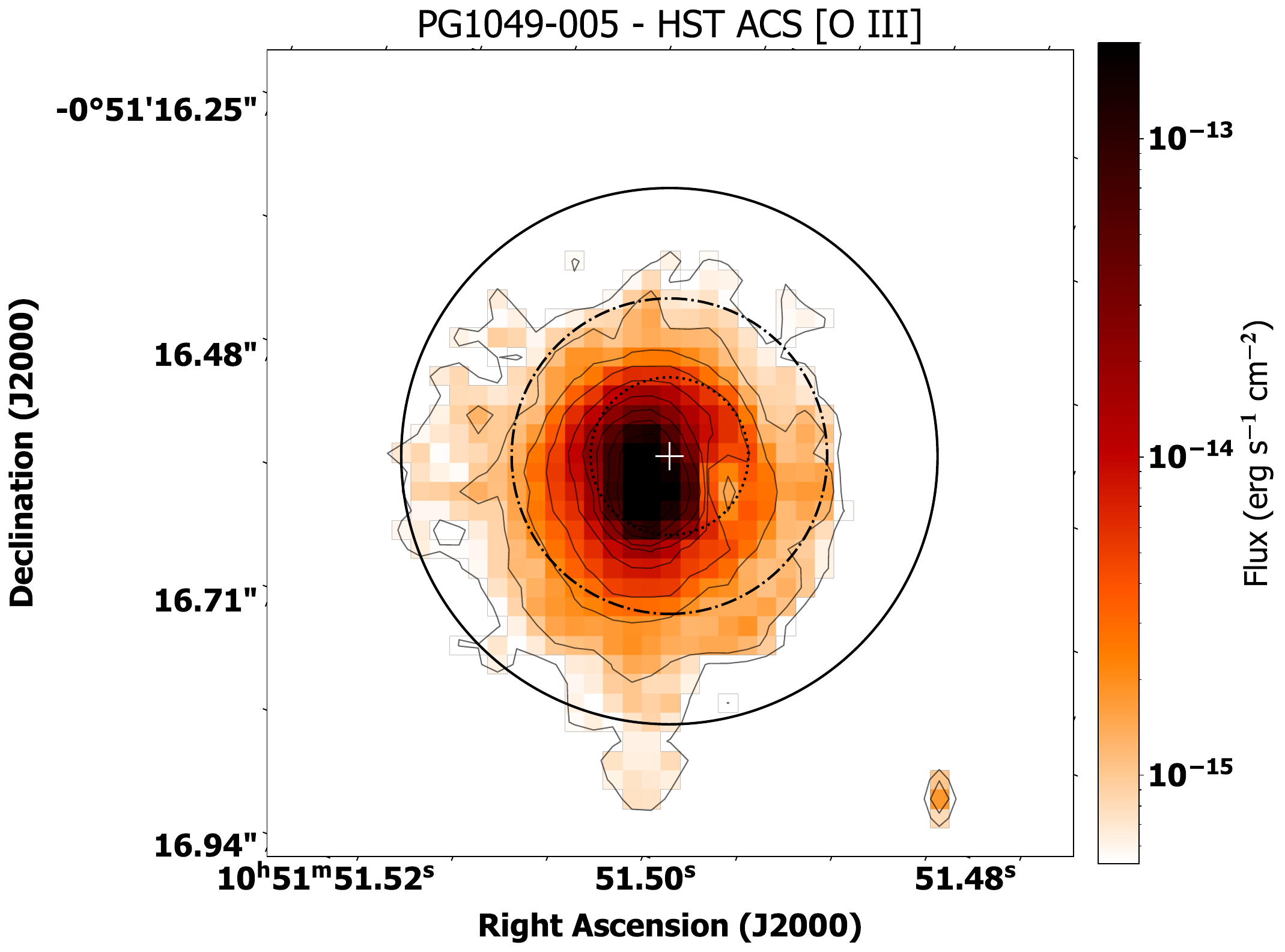}
     \end{subfigure}
     
     \centering
     \begin{subfigure}{0.49\textwidth}
         \centering
        \includegraphics[width=\linewidth]{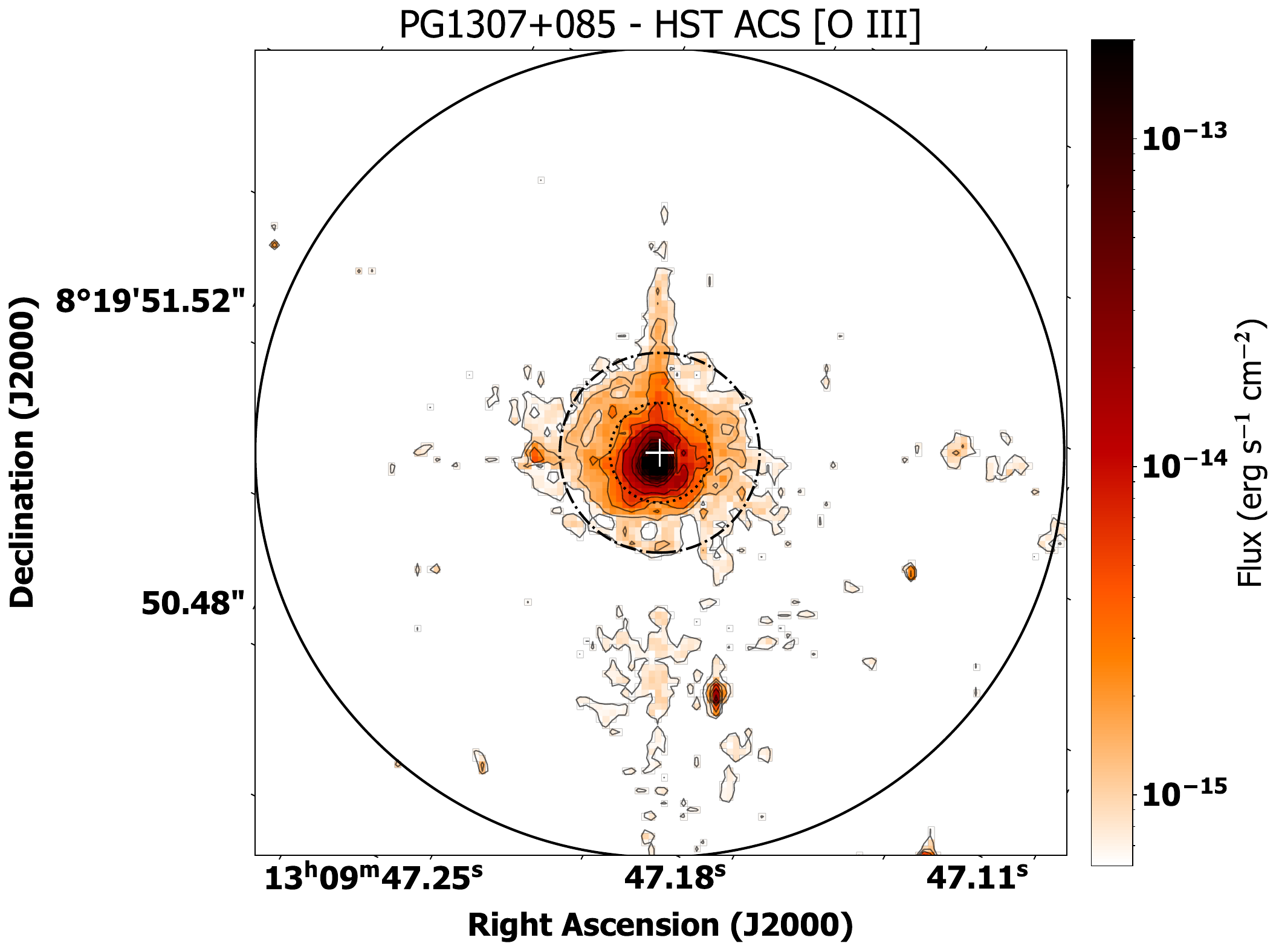}
     \end{subfigure}

     \caption{\textit{continue}}
     \label{fig:morphology2}
\end{figure*}

All galaxies show extended [O~III] emission, up to several kpc in the plane of the sky. An elongated structure resembling ionisation cones is clearly observed in one target, PG0157+001 (Fig. \ref{fig:morphology1}), while the remaining targets show a more "face-on" morphology. We measure the total projected NLR extent of the ionised emission in the [O~III] continuum-subtracted images as the distance between individual continuum centroids (white crosses) and the farthest radial extent in each image, as determined by 3$\sigma_{\rm sky}$ flux contours (excluding any contribution from diffraction spikes). Table \ref{tab:qso1s} lists the measured NLR sizes, $R_{\rm [O~III]}$, and 3$\sigma_{\rm sky}$ values for each target. 

\begin{table*}
\begin{center}
\begin{tabular}{cccccc}
\multicolumn{1}{c}{}\\
\hline
\multicolumn{1}{l}{QSO1s}
&\multicolumn{1}{c}{$F_{\rm[O~III]}$}
&\multicolumn{1}{c}{$L_{\rm[O~III]}$}
&\multicolumn{1}{c}{$R_{\rm [O~III]}$}
&\multicolumn{1}{c}{3$\sigma_{\rm sky}$ Flux}
&\multicolumn{1}{c}{Asymmetry}

\\
& (erg~s$^{-1}$~cm$^{-2}$)& (erg~s$^{-1}$)&(kpc) &(erg~s$^{-1}$~cm$^{-2}$) & ($b/a$)\\
\hline
\hline
  PG0026+129  &7.7$^{+0.1}_{-0.1}$$\times$10$^{-14}$&3.7$^{+0.1}_{-0.1}\times10^{42}$& 4.7$^{+0.1}_{-0.1}$ & 4.4$\times10^{-18}$&1.4\\
  
  PG0052+251 &6.9$^{+0.1}_{-0.1}$$\times$10$^{-14}$& 3.8$^{+0.1}_{-0.1}$$\times10^{42}$ & 5.0$^{+0.6}_{-0.3}$ &4.7$\times10^{-18}$& 2.4\\
  
  PG0157+001  &4.6$^{+0.1}_{-0.1}\times$10$^{-14}$&2.8$^{+0.1}_{-0.1}\times10^{42}$ &9.1$^{+0.3}_{-0.3}$ &5.9$\times10^{-18}$& 5.6\\
  
  PG0953+414  & 7.9$^{+0.1}_{-0.1}$$\times$10$^{-14}$&1.0$^{+0.1}_{-0.1}$$\times10^{43}$&3.2$^{+0.4}_{-0.4}$ & 3.8$\times10^{-18}$&1.5\\
  
  PG1012+008 & 3.0$^{+0.1}_{-0.1}$$\times$10$^{-14}$&2.5$^{+0.1}_{-0.1}$$\times10^{42}$ & 3.0$^{+0.3}_{-0.4}$ &5.4$\times10^{-18}$&1.2\\
  
  PG1049\textminus005 &3.5$^{+0.1}_{-0.1}$$\times$10$^{-14}$& 1.1$^{+0.1}_{-0.1}$$\times10^{43}$ &3.4$^{+0.5}_{-0.5}$ & 2.9$\times10^{-18}$&1.2\\
  
  PG1307+085  &6.9$^{+0.1}_{-0.1}\times$10$^{-14}$&3.9$^{+0.1}_{-0.1}\times10^{42}$&8.1$^{+0.3}_{-2.9}$ & 5.1$\times10^{-18}$&3.1\\

\hline
\end{tabular}
\end{center}
\caption{\textbf{Imaging measurements for the PG QSO1s from this work.} Note: (1) Galaxy name; (2) [O~III] integrated flux measured in the \textit{HST} image within apertures encompassing the full extent of the optical NLR; (3) total [O III] 5007\AA~luminosity, integrated above 3$\sigma_{\rm sky}$; (4) NLR radius; (5) 3$\sigma$ standard deviation of the sky value in the [O~III] image, as used for our measurements of $R_{\rm [O~III]}$ and $L_{\rm[O~III]}$; (5) NLR asymmetry measurement.}
\label{tab:qso1s} 
\end{table*}

The uncertainties in the measurements are estimated as the difference between the values measured using contours at 2$\sigma_{\rm sky}$ and 4$\sigma_{\rm sky}$, following \citet{storchi2018a}. Among the seven quasars in the sample, four objects (PG0026+129, PG0953+414, PG1012+008, and PG1049\textminus005) show relatively symmetric morphologies, with $b/a$ between 1.2-1.5, where $b$ reflects the major axis (maximum measured radius, $R_{\rm [O~III]}$), and $a$ is the minor axis (minimum measured radius measured in the perpendicular direction). The remaining three targets (PG0052+251, PG0157+001, and PG1307+085) display more asymmetric, extended [O~III] emission, with $b/a$ in the 2.4-5.6 range. Individual $b/a$ values are listed in Table \ref{tab:qso1s}.

We note that our measurements of NLR radii do not take into account any emission contribution from CTE (charge transfer efficiency), and/or point-spread function (PSF), which in turn do not affect our values of $R_{\rm [O~III]}$. In the case of CTE, electrons from the pixels corresponding to the nuclear PSF get trapped when moved in the instrument readout, creating a long streak, usually to the top of the image. These features have low surface brightness, and do not contribute to our measurements of the extended [O~III] emission. Meanwhile, performing a very accurate PSF subtraction is not possible for the available data. However, given that the continuum and line images used in this work were observed with similar wavelengths, one should in principle be able to achieve a very good PSF subtraction by just subtracting the continuum from the emission line image, but residuals are always expected. These residuals are taken into account when performing our measurements, and the affected regions are disregarded from the final NLR size values.

\subsection{Ionised Gas Luminosity}
\label{sec:luminosities}
The total [O~III] luminosities, $L_{\rm [O III]}$, as derived from the continuum-subtracted images (see Section \ref{sec:data_reduction}) are obtained by integrating individual fluxes values above 3$\sigma_{\rm sky}$, and the distances listed in Table \ref{tab:targets}. The resulting values are listed in Table \ref{tab:qso1s}.

\begin{table*}
\begin{center}
\begin{tabular}{ccccccc}
\multicolumn{5}{c}{}\\
\hline
\multicolumn{1}{l}{QSO2s}
&\multicolumn{1}{c}{$L_{\rm [O~III]}$}
&\multicolumn{1}{c}{$Q_{\rm [O~III]}$}
&\multicolumn{1}{c}{$R_{\rm [O~III]}$}
&\multicolumn{1}{c}{$M_{\rm[O~III]}$}
&\multicolumn{1}{c}{$R_{\rm max [O~III]}$}
&\multicolumn{1}{c}{\ul{$R_{\rm [O~III]}$}}
\\
&(erg~s$^{-1}$)&(photons s$^{-1}$)&(kpc)&(10$^{5}M_{\odot}$)&(kpc)&$R_{\rm max [O~III]}$\\
\hline
\hline
SDSS J11524 & $3.5\times10^{42}$&$4.8\times 10^{54}$ & 3.6 & 0.17& 3.6 & 0.99 \\

B2 1435  &$7.1\times10^{42}$& $9.6\times 10^{54}$& 3.8 & 0.04& 5.1 & 0.75\\

2MASX J07594$^{\dagger}$ &$8.6\times10^{42}$& $1.2\times 10^{55}$& 3.2 & 4.44& 5.7 & 0.57 \\

2MASX J08025  & $8.0\times10^{42}$& $1.1\times 10^{55}$& 3.6 & 11.8& 5.4 & 0.67\\

MRK 34  & $5.7\times10^{42}$& $7.8\times 10^{54}$& 2.2 & 337.0& 4.3 & 0.51 \\

2MASX J11001$^{\dagger}$  &$8.5\times10^{42}$&  $1.2\times 10^{55}$& 3.0 & 25.6& 4.7 & 0.64\\

FIRST J12004  & $1.3\times10^{43}$& $1.7\times 10^{55}$& 6.1 & 140.0& 6.8 & 0.89  \\

2MASX J14054$^{\dagger}$  & $4.6\times10^{42}$& $6.2\times 10^{54}$& 1.0 & 8.94& 4.1 & 0.25\\

MRK 477$^{\dagger}$ &  $4.0\times10^{42}$& $5.4\times 10^{54}$& 2.7 & 152.0& 3.8 & 0.72\\

2MASX J16531  & $1.1\times10^{43}$& $1.5\times 10^{55}$& 6.5 & 2.54& 6.4 & 1.01\\

2MASX J17135$^{\dagger}$  & $6.1\times10^{42}$& $8.3\times 10^{54}$ & 1.8& 24.4& 4.7 & 0.38\\

2MASX J13003  & $4.8\times10^{42}$& $6.5\times 10^{54}$& 4.7 & 0.11& 4.2 & 1.11\\

\hline
\end{tabular}
\end{center}
\caption{\textbf{Imaging measurements of the QSO2s in \citetalias{trindadefalcao2021a}.} Note: (1) Galaxy name; (2) total [O III] 5007\AA~luminosity, integrated above 3$\sigma_{\rm sky}$; (3) number of ionising photons per second emitted by the AGN; (4) measured NLR radius; (5) total ionised gas mass; (6) maximum NLR radius estimated using $L_{\rm [O~III]}$; (7) ratio between measured NLR size and estimated maximum NLR size. Targets marked with a \textbf{$^{\dagger}$} are described as compact sources by \citet{fischer2018a}.}
\label{tab:qso2s} 
\end{table*}

\subsection{Individual Targets}
\label{sec:individual targets}

In this section we summarize our main findings concerning the NLR morphologies of individual targets. 
\medskip

\textbf{PG0026+129:} This target displays a symmetric ($b/a$=1.4) NLR, with $R_{\rm [O~III]}$=4.7~kpc, and $L_{\rm [O~III]}$=3.7$\times10^{42}$~erg~s$^{-1}$. The extended [O~III] emission (Fig. \ref{fig:morphology1}) shows two elongated and narrow structures to the south and west of the nucleus, possibly due to diffraction spikes. An elliptical [O~III] structure is detected within the south "arm".

\smallskip
\textbf{PG0052+251:} The [O~III] NLR in PG0052+251 has a measured radial extent of $R_{\rm [O~III]}$=5.0~kpc, with a luminosity $L_{\rm [O~III]}$=3.8$\times10^{42}$~erg~s$^{-1}$, and $b/a$=2.4. As shown in Fig. \ref{fig:morphology1}, three "blob-like" structures are observed to the east of the nucleus, embedded in a region of lower surface brightness. The farthest blob is located $\sim$4.3~kpc from the AGN. Diffraction spikes are also observed in this target, in this case, extending to the north of the nuclear source. 

\smallskip
\textbf{PG0157+001:} Among the PG QSOs in our sample, PG0157+001 has the most extended optical NLR, $R_{\rm [O~III]}$=9.1~kpc, with $L_{\rm [O~III]}$=2.8$\times10^{42}$~erg~s$^{-1}$, and $b/a$=5.6. The bulk of ionised emission is located northeast of the nucleus, where a region of enhanced [O~III] emission is observed $\sim$2.8~kpc from the AGN (Fig. \ref{fig:morphology1}). At larger radii, we observe an "L-shaped" structure located $\sim$9~kpc from the nuclear source. A VLA-A 8.4-GHz radio map of PG0157+001 shows a triple radio source extending over $\sim$6.8 kpc, and oriented along the same NE-SW direction \citep{leipski2006a}, and clearly related to the optical features shown in Fig. \ref{fig:morphology1}. 

\smallskip
\textbf{PG0953+414:} This target has a NLR with $R_{\rm [O~III]}$=3.2~kpc, $L_{\rm [O~III]}$=1.0$\times10^{43}$~erg~s$^{-1}$, and $b/a$=1.5. In Fig. \ref{fig:morphology1}, the elongated and narrow structure observed to the north of the central source is attributed to diffraction spikes. 

\smallskip
\textbf{PG1012+008:} PG1012+008 is part of an interacting system \citep{bahcall1997a}, with a neighboring galaxy located $\sim$10.5~kpc from the nucleus (not shown). The [O~III] NLR has a radius of $R_{\rm [O~III]}$=3.0 kpc, showing a somewhat "pear-shaped" morphology ($b/a$=1.2) (Fig. \ref{fig:morphology2}), and a luminosity $L_{\rm [O~III]}$=2.5$\times10^{42}$~erg~s$^{-1}$. We also observe [O~III] diffraction spikes extending to the north of the nucleus in this object. 

\smallskip
\textbf{PG1049\textminus005:} This galaxy has the highest redshift among the QSO1s analyzed in this work, $z$=0.360. The optical NLR has a radius of $R_{\rm [O~III]}$=3.4~kpc (Fig. \ref{fig:morphology2}), $b/a$=1.2, with $L_{\rm [O~III]}$=1.1$\times10^{43}$~erg~s$^{-1}$, showing diffraction spikes to the south of the AGN. 

\smallskip
\textbf{PG1307+085:} This target features an NLR with a radius $R_{\rm [O~III]}$=8.1~kpc, making it the second most extended NLR in our sample, with $L_{\rm [O~III]}$=3.9$\times10^{42}$~erg~s$^{-1}$, and $b/a$=3.1. The continuum-subtracted [O~III] image (Fig. \ref{fig:morphology2}) reveals the presence of diffraction spikes to the north of the AGN, and the bulk of the extended emission is observed to the south of the nuclear source.

\bigskip

Among the QSO1s in our sample, PG0157+001 is the only object that shows a clear spatial correlation between optical and radio \citep{leipski2006a} NLR morphologies. \citet{fischer2021a} suggests that extended radio structures observed in radio-quiet AGNs may arise from interactions between AGN-driven winds and dense dust lanes within the galaxy. In this scenario, the winds may not efficiently displace the dense dust lanes, leading to shock generation and the production of synchrotron emission as a byproduct. Enhanced synchrotron radiation may also arise from relativistic particles through cosmic ray accelerations at the shock front \citep[][]{fischer2021a}. Thus, while the radio structure observed in the 8.4 GHz map of PG0157+001 by \citet{leipski2006a} could potentially be a jet, it is equally plausible it is associated with its distinct and extended NLR (Fig. \ref{fig:morphology1}), a characteristic more commonly associated with type 2 AGNs.

The seven PG QSO1s in our sample were previously observed by \citet{bennert2002a} with the Wide Field Planetary Camera 2 (WFPC2) on board \textit{HST}. While PG0026+139 was observed with the Planetary Camera (PC) with a resolution of 0.0455$''$, the other six targets were imaged with the Wide Field Camera (WFC) at a resolution of 0.1$''$. However, the dataset used by \citet{bennert2002a} lacked a consistent set of continuum images for on-band subtraction. Specifically, useful continuum images were only available for two targets in the sample, PG0157+001, and PG0953+414. For PG1049-005, the continuum image was contaminated by line emission, for which they corrected by scaling the continuum. For the other four targets, \citet{bennert2002a} used a scaled stellar PSF for continuum subtraction instead.

Comparing our results to those of \citet{bennert2002a} shows that, although they did detect extended [O~III] emission in all QSO1s in the sample, their individual $R_{\rm [O~III]}$ values are not generally consistent with the results from this study shown in Table \ref{tab:qso1s}. However, a direct comparison between the results from this work and those of \citet{bennert2002a} can only be performed for 2 sources, PG0052+251, and PG0157+001, given that these are the only two objects for which they published their images. For PG0052+251, \citet{bennert2002a} found $R_{\rm [O~III]}=5.6^{+0.8}_{-0.8}$~kpc, while this work finds $R_{\rm [O~III]}=5.0^{+0.6}_{-0.3}$~kpc, and thus the numbers are consistent within the uncertainties. In the case of PG0157+001, \citet{bennert2002a} measured a NLR extent of $R_{\rm [O~III]}=5.3^{+0.9}_{-0.9}$~kpc, while this work measures $R_{\rm [O~III]}=9.1^{+0.3}_{-0.3}$~kpc, a factor of $\sim$1.7 larger. We attribute this difference to the difference in sensitivity between ACS and the WFPC2/WFC\footnote{https://hst-docs.stsci.edu/acsihb/chapter-5-imaging/5-2-important-considerations-for-acs-imaging\#id-5.2ImportantConsiderationsforACSImaging-5.2.35.2.3LimitingMagnitudes}. 

For the remaining five sources, we find a larger NLR extent than that of \citet{bennert2002a} for two sources, PG0026+129 ($\sim$1.6 times larger), and PG1307+085 ($\sim$2.0 times larger), which we attribute to the higher sensitivity of ACS compared to WFPC2/WFC. This is illustrated by the large lower limit of uncertainty in $R_{\rm [O~III]}$ for PG1307+085 (-2.9~kpc, Table \ref{tab:qso1s}), which was calculated considering values measured at 2$\sigma_{\rm sky}$ (Section \ref{sec:morphology}). For PG0953+414, PG1012+008, and PG1049-005, however, the NLR sizes in \citet{bennert2002a} are $\sim$2-3 times larger than the ones measured in this work. We note, however, that visual inspection of the \textit{HST}/ACS images used in this work, both before and after continuum subtraction, indicates that there is no line emission at the scales reported by \citet{bennert2002a}, which were measured at radii of 1.5$''$, 1.6$''$, and 2.0$''$ for PG0953+414, PG1012+008, and PG1049-005, respectively. Given that these images were not published in their manuscript, we cannot confirm what may have caused such a difference.

\section{Discussion}
\label{sec:discussion}

\subsection{Relation Between $R_{\rm [O~III]}$ and $L_{\rm [O~III]}$}
\label{sec:relation_r_and_L}

The relation between the size of the extended NLR ($R_{\rm [O~III]}$) and its [O~III] luminosity ($L_{\rm [O~III]}$) has been analysed in several studies \citep[e.g.,][]{bennert2002a, schmitt2003a, schmitt2003b, storchi2018a, fischer2018a}. \citet{schmitt2003a,schmitt2003b} used \textit{HST} observations of the [O~III] emission in 60 low-luminosity ($L_{\rm [O~III]}<10^{42}$~ergs~${\rm s^{-1}}$) AGNs and found that $R_{\rm [O~III]}$ scales with $L_{\rm [O~III]}$ to the power of 0.33. This was interpreted as consistent with photoionisation and a roughly constant density throughout the entire NLR, analogous to a Strömgren sphere - an ionised region where the rate of ionisation balances the rate of recombination, resulting in a radius that depends on the luminosity of the ionising source. This suggests that the size of the ionised region increases with AGN luminosity, similar to the way the Strömgren radius depends on the ionising source in H~II regions.

Ground-based studies of QSO2 NLR sizes \citep[e.g.,][]{greene2011a, liu2013a} have also explored the relationship between NLR radius and luminosity. These studies, utilizing data from imaging and long-slit spectroscopy, found a scaling exponent of $\gamma$=0.22. This suggests that the size of the NLR may be limited by ionisation state and density rather than the number of ionising photons (i.e., a matter-bounded scenario). Note that \citet{greene2011a} and \citet{liu2013a} used different observational setups, potentially affecting the accuracy of their measurements. While \citet{greene2011a} employed a 1$''$ $\times$ 4$'$ slit at the Magellan/Clay telescope, with a seeing of approximately 1$''$ over the two runs, \citet{liu2013a} used GMOS-N IFU (seeing $<$0.75$''$ with IQ=70\% or IQ=20\%, or 1.5$''$ with IQ=any\footnote{https://www.gemini.edu/instrumentation/gmos/capability}) with a two-slit mode covering a 5$\times$7 arcsec$^{2}$ field of view. In both cases, the accuracy of their results was affected by the size of their slits and the seeing conditions, which are not limitations with the \textit{HST} imaging used in this work.

Recently, \citet{storchi2018a} used \textit{HST}/ACS [O~III] imaging to analyse the NLR of 9 luminous ($L_{\rm [O~III]}>$10$^{42}$~ergs~${\rm s^{-1}}$) QSO2s (0.1$<$\textit{z}$<$0.5). Combining their results of $R_{\rm [O~III]}$ and $L_{\rm [O~III]}$ with those from previous studies targeting lower redshift and lower luminosity AGNs \citep{bennert2002a, schmitt2003b, fischer2018a}, they obtained three linear relations, varying with AGN type. For type 1 AGNs they found a slope of $\gamma_{1}$=0.57$\pm$0.05; for type 2 AGNs, $\gamma_{2}$=0.48$\pm$0.03; and considering all AGNs, $\gamma_{\rm all}$=0.51$\pm$0.03.
 
\begin{figure*}
    \centering
    \includegraphics[width=0.7\textwidth]{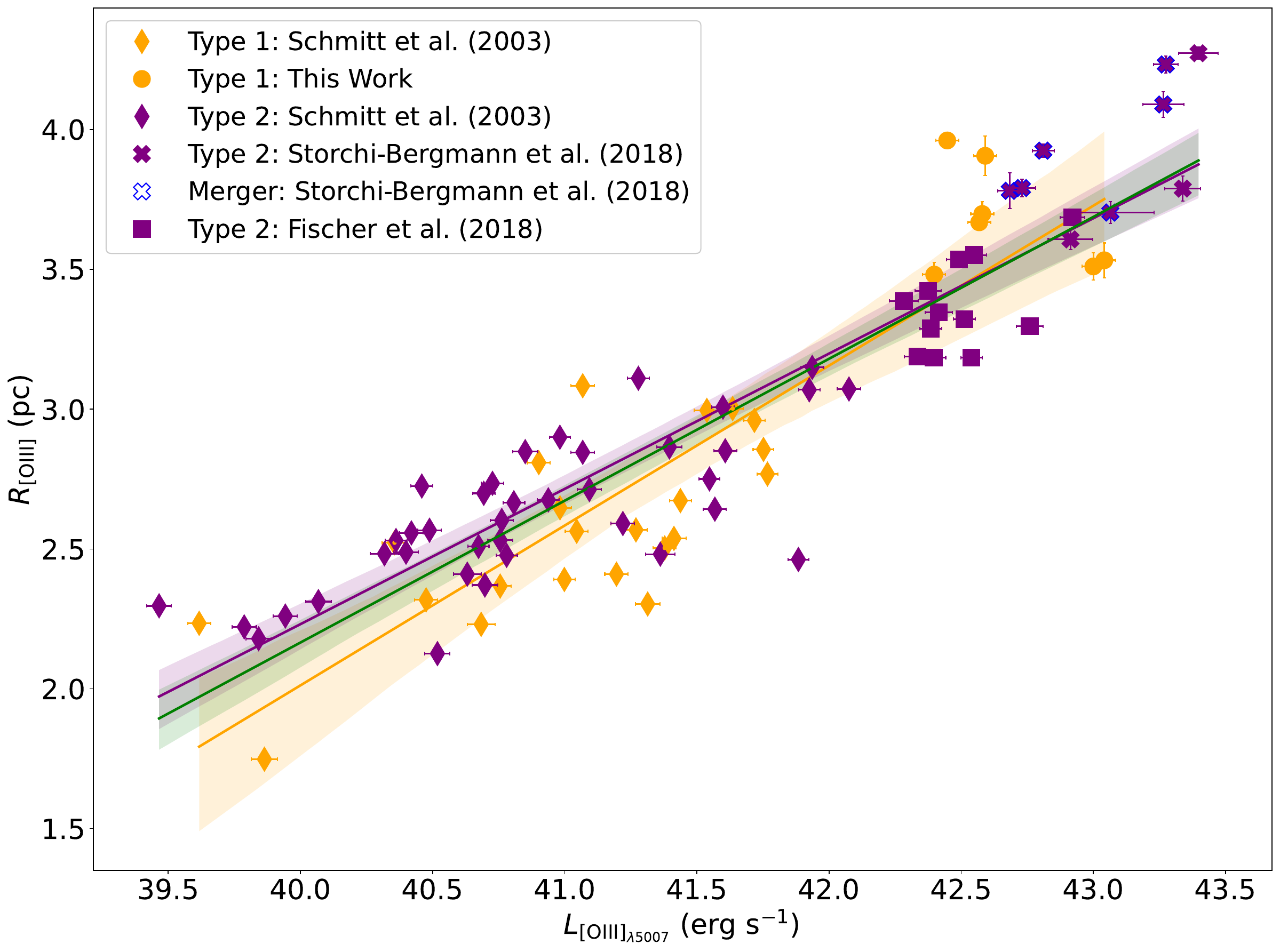}
    \caption{Relation between the extent of the NLR and its [O~III] luminosity, combining the measurements from this study (orange circles), along with the results of \citet{schmitt2003b} (orange and purple diamonds), \citet{storchi2018a} (purple X's), and \citet{fischer2018a} (purple squares). We also show the results of the linear regressions for each type of AGN (see text), with 1$\sigma$ confidence intervals shown as shaded regions. The AGNs in \citet{storchi2018a} that appear to be in interaction are highlighted with blue contours. Note the larger $R_{\rm [O~III]}$ values for the QSO1s (this work) compared to QSO2s of similar $L_{\rm [O~III]}$ \citep{fischer2018a}.}
    \label{fig:comparison}
\end{figure*}

With our results, we can now revisit the observed $R_{\rm [O~III]}$-$L_{\rm [OIII]}$ relation between NLR size and [O~III] luminosity. Our results are compared to the results from \citet{schmitt2003b}, \citet{fischer2018a}, and \citet{storchi2018a}. The new relation is shown in Fig. \ref{fig:comparison}, in which different colors differentiate type 1 (orange) and type 2 (purple) AGNs, and different samples are plotted with different markers. 

The size of the NLR increases with increasing luminosity, until the maximum luminosity considered, $L_{\rm [OIII]}=10^{43.3}$~ erg~s$^{-1}$. As in \citet{storchi2018a}, we apply a linear least-squares regression to the data, and find three linear relations to account for variations in AGN type. The first relation is restricted to type 1 AGNs:
\begin{equation}
\label{eq:trindade_relation1}
     {\rm log} (R_{\rm [O~III]})_{1} = (0.57\pm 0.05){\rm log}(L_{\rm [O~III]}) -20.86\pm2.35
\end{equation}

\noindent shown as an orange line in Fig. \ref{fig:comparison}. The second relation is for type 2 AGNs:
\begin{equation}
\label{eq:trindade_relation2}
    {\rm log} (R_{\rm [O~III]})_{2} = (0.48\pm0.02){\rm log}(L_{\rm [O~III]}) -17.13\pm1.07
\end{equation}

\noindent shown as a purple line in Fig. \ref{fig:comparison}. The third relation considers
all AGNs:
\begin{equation}
\label{eq:trindade_relation_all}
    {\rm log} (R_{\rm [O~III]})_{\rm all} = (0.51\pm0.03){\rm log}(L_{\rm [O~III]}) -18.13\pm1.06
\end{equation}
\noindent shown as a green line in Fig. \ref{fig:comparison}. We show the 1$\sigma$ confidence interval of each linear regression fit as shaded regions, calculated using the bootstrap method to estimate confidence intervals \footnote{ https://seaborn.pydata.org/generated/seaborn.regplot.html}.

The slope found in Equation \ref{eq:trindade_relation_all} for all AGNs ($\gamma_{\rm all}$=0.51) is consistent with photoionisation, as we demonstrate in Section \ref{sec:photoionisation_molecular_clouds}. When considered separately, type 1s show a steeper dependency of the NLR (Equation \ref{eq:trindade_relation1}) extent on the [O~III] luminosity than type 2s (Equation \ref{eq:trindade_relation2}), with overall larger NLRs. We revisit this point in Section \ref{sec:Constraints_on_NLR_Sizes}.

\subsubsection{Photoionisation of ISM clouds}
\label{sec:photoionisation_molecular_clouds}

The ionisation state of a slab of gas can be generally described using the dimensionless ionisation parameter, $U$:

\begin{equation}
\label{eq:ioniz_param_photoionization}
U=\frac{Q}{4\pi\textit{c}~n_{\rm H}r^{2}}
\end{equation}

\noindent where $Q$ is the number of ionising photons emitted by the AGN per second, $r$ is the distance to the AGN in cm, $n_{\rm H}$ is the hydrogen number density in cm$^{-3}$, and $c$ is the speed of light in cm~s$^{-1}$. Rearranging this equation, we obtain:

\begin{equation}
\label{eq:radius_photoionization}
r = \sqrt{\frac{Q}{4\pi\textit{c}~ n_{\rm H}U}}
\end{equation}

\citetalias{trindadefalcao2021a} shows that the number of ionising photons per second (\textit{Q}) is proportional to the bolometric luminosity ($L_{\rm bol}$) of the AGN, when a consistent SED is used (see Section \ref{sec:nlr_sizes_t1_t2}). It also shows that $L_{\rm [O~III]}$ can be used to estimate $L_{\rm bol}$, after applying a bolometric correction factor \citep{lamastra2009a}. Thus, from \citetalias{trindadefalcao2021a}, $L_{\rm [O~III]}$ is also proportional to $Q_{\rm [O~III]}$. We note that there is a considerable debate surrounding the proper bolometric luminosity correction factor that may be applied to $L_{\rm [O~III]}$ to obtain $L_{\rm bol}$, including, but not limited to, the role of internal extinction, and whether such a correction is luminosity-dependent \citep[e.g.,][]{netzer2006a, lamastra2009a, stern2012a}. We further discuss this point in the context of the PG QSOs in this work in Section \ref{sec:bolometric_luminosity}.

For photoionised gas in the ISM, Equation \ref{eq:radius_photoionization} shows that the radial distance $r$ scales with the square root of \textit{Q}, for a gas characterised by a specific combination of density and ionisation parameter. This suggests a direct relationship between NLR sizes and [O~III] luminosity: $R_{\rm [O~III]}\sim L_{\rm [O~III]}^{0.5}$, consistent with the slope found for all AGN in Equation \ref{eq:trindade_relation_all}. Note that this approach solely applies a correction factor to $L_{\rm [O~III]}$ to obtain $L^{\rm [O~III]}_{\rm bol}$ (and thus $Q_{\rm [O~III]}$), and does not rely on specific gas properties, such as density ($n_{\rm H}$), or ionisation parameter (\textit{U}). In other words, we focus on the direct relationship between $r$-\textit{Q} in Equation \ref{eq:radius_photoionization}, rather than calculating the actual size of the ionised regions, for which we would need constraints on \textit{U} and $n_{\rm H}(r)$. 



A size-luminosity slope of 0.5 was also identified by \citet{bennert2002a} when examining a sample of quasars (the PG QSO1s in this study) and Seyfert 2 galaxies \citep{falcke1998a}. Their analysis covered a range of 3 orders of magnitude in luminosity. As \citet{bennert2002a} explained, this slope is anticipated if all AGNs share, on average, the same density, ionisation parameter, and SED. This suggests a self-regulating mechanism that governs the size of the NLR to scale with $U$, expressed as $U\propto Q/(R^{2}_{\rm [O~III]}n_{\rm H})$, with $Q\propto L_{\rm [O~III]}$, resulting in $R_{\rm [O~III]}\propto L_{\rm [O~III]}^{0.5}$, for a given $n_{\rm H}$ and $U$. 

The relationship between NLR size and luminosity observed here and in \citet{bennert2002a} indicates that all AGNs possess a similar value of ($n_{\rm H}\times U$) at the boundary of their NLR. This results in efficient [O~III] emission originating from regions where $n_{\rm H}\times U\sim0.1$\footnote{\citet{bennert2002a} defines the NLR outskirts as having $n_{\rm H}\sim10^{2-3}$~cm$^{-3}$ and $U\sim10^{-(2-3)}$}. This is independent of the specific spatial values of $U$ and $n_{\rm H}$.

\subsection{Constraints on NLR Sizes}
\label{sec:Constraints_on_NLR_Sizes} 

Now, we analyse whether Equation \ref{eq:radius_photoionization} can be used to estimate a ``maximum" NLR size, $R_{\rm max}$, when applied constraints on the density and ionisation state of the gas. 

Specifically for [O~III] 5007\AA~, \citet{revalski2022a} have shown that production of [O~III] gas occurs within a limited range in ionisation parameter. While the strongest emission arises from gas with \textit{U}$\sim10^{-2}$, there can be contributions to [O~III] from lower-ionisation gas. Therefore, here we consider \textit{U}$\sim10^{-3}$ as the "lower limit" for production of [O~III]-emitting gas. Meanwhile, for the gas density, \citet{fischer2017a} have shown that much of the gas found in the NLR originates from the dissociation and photoionisation of molecular cloud reservoirs. These typically have densities $n_{\rm H}\sim$ 100 cm$^{-3}$ \citep[e.g.,][]{peterson1997a, lequeux2005a}. Therefore, $R_{\rm max}$ (in cm) can be estimated as:  

\begin{equation}
\label{eq:r_max}
 R_{\rm max}\sim \sqrt{\frac{\textit{Q}}{4\pi\textit{c}(10^{2})(10^{-3})}}\sim \sqrt{\frac{\textit{Q}}{3.8\times10^{10}}}
\end{equation}

Physically, Equation \ref{eq:r_max} gives the maximum radial distance at which the ISM can be sufficiently ionised to emit [O~III] 5007\AA, which scales with $\sim Q^{0.5}$. In Section \ref{sec:nlr_sizes_t1_t2}, we test the accuracy of Equation \ref{eq:r_max} in predicting the sizes of the NLR by applying it to the QSO2s in \citetalias{trindadefalcao2021a}, and to the PG QSO1s studied here. 

\subsubsection{NLR Sizes in type 1 \& type 2 AGNs}
\label{sec:nlr_sizes_t1_t2}

In \citetalias{trindadefalcao2021a}, we used the observed \textit{NuSTAR} X-ray luminosity for the QSO2 Mrk 34 \citep{gandhi2014a} to estimate \textit{Q} for this AGN (Table \ref{tab:qso2s}). We assumed an SED described by a number of broken power-laws of the form $L_{\nu} \sim \nu^{-\alpha}$, where $\alpha$, the spectral or energy index, is a positive number \citep[e.g.,][]{melendez2011a}. Specifically, our models used:

\begin{align*}
\alpha = \left\lbrace
\begin{array}{r@{}l}
    1.0, \qquad & \qquad \qquad h\nu < 13.6~{\rm eV} \\
    1.4, \qquad & 13.6~{\rm eV} \leq {h\nu} \leq 0.5~{\rm keV} \\
    0.5, \qquad & 0.5~{\rm keV} \leq {h\nu} \leq 100~{\rm keV}
\end{array}
\right.
\end{align*}

The derived $Q$/$L_{\rm [O~III]}$ ratio for Mrk 34 was then used to estimate $Q_{\rm [O~III]}$ for the remaining QSO2s in the sample (Table \ref{tab:qso2s}). 

\bigskip

Now, we use the values of $Q_{\rm [O~III]}$ from \citetalias{trindadefalcao2021a} (listed in Table \ref{tab:qso2s}) to estimate $R^{\rm [O~III]}_{\rm max}$ for the QSO2s, according to Equation \ref{eq:r_max}. However, there is a caveat: \citet{fischer2018a} divided the QSO2s in \citetalias{trindadefalcao2021a} into two groups based on [O~III] NLR morphology. 
\begin{description}
    \item \textbf{Compact sources:} These tend to have broader nuclear FWHMs ($>$1500 km s$^{-1}$).
    \item \textbf{Extended sources:} These objects have narrower nuclear FWHMs ($<$900 km s$^{-1}$). 
\end{description}

\begin{figure}
    \centering
    \includegraphics[width=.5\textwidth]{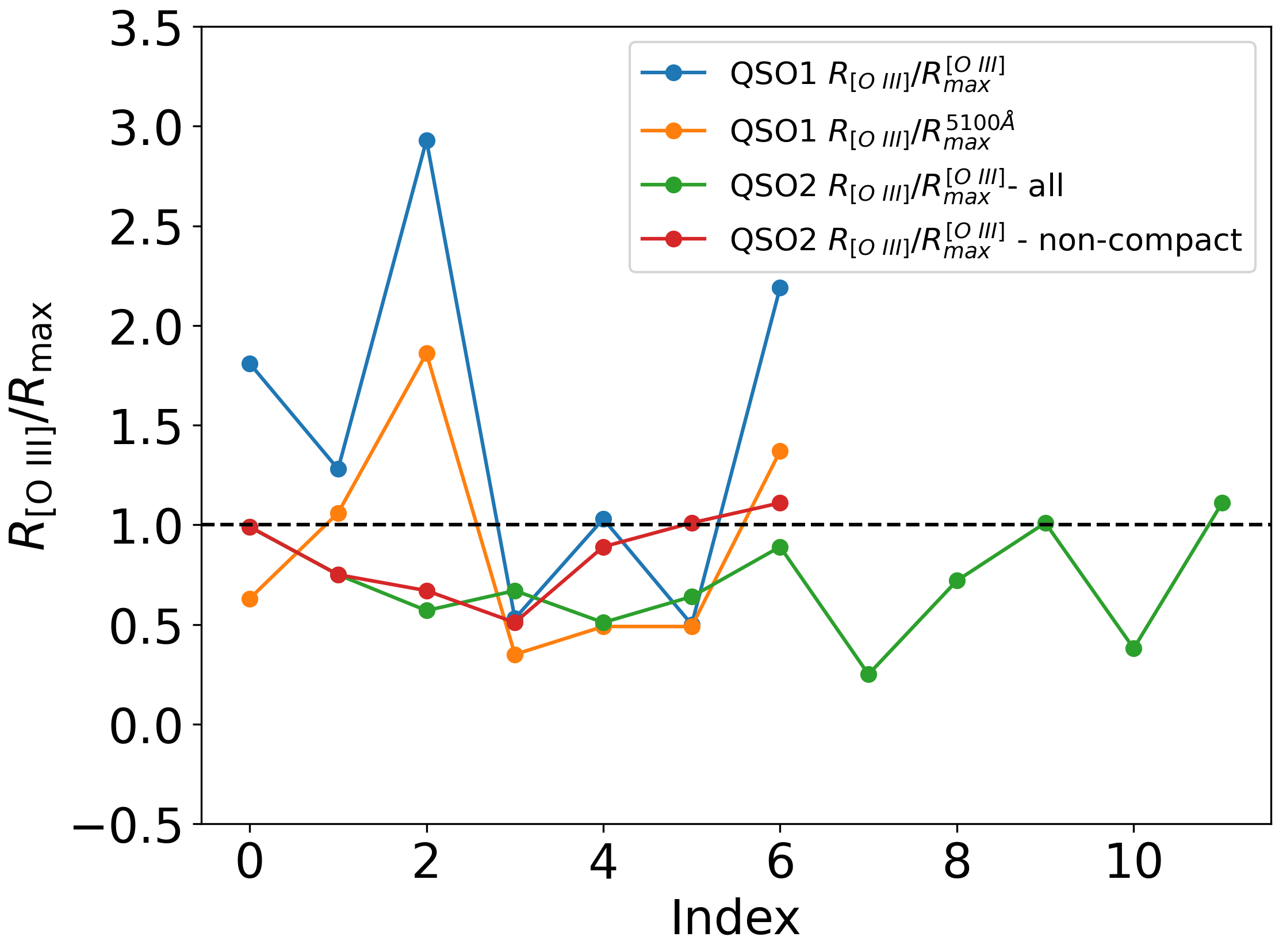}
    \caption{Deviation plot for the different $R_{\rm [O~III]}/R_{\rm max}$ ratios for the PG QSO1s, and QSO2s. The x-axis represents the index, which corresponds to the position of each data point within the dataset (i.e., the order in which the values appear in the dataset). The y-axis indicates how far each value deviates from $R_{\rm [O~III]}/R_{\rm max}$=1 (Tables \ref{tab:qso2s}, \ref{tab:qso1s_rmax}, \ref{tab:qso1s_lbol_cont}). These reveal differences in the $R^{\rm  [O~III]}_{\rm max}$ distributions between the samples. Specifically, using [O~III] to estimate the NLR sizes results in significantly larger (overestimated) sizes for QSO1s (in blue). Estimating the bolometric luminosity from the 5100\AA~continuum luminosity (in orange) yields a more similar distribution between the samples. We show $R_{\rm [O~III]}/R_{\rm max}$=1 as a black dashed line, for comparison.}
    \label{fig:ideal}
\end{figure}

\citet{fischer2018a} propose that the differences in NLR morphology observed in the QSO2s may be caused by more gas near the AGN in compact targets, which would attenuate the ionising radiation and prevent it from reaching further into the NLR of these targets, thus leading to smaller NLR sizes. Therefore, to compare the measured $R_{\rm [O~III]}$ (Table \ref{tab:qso2s}) to those estimated using Equation \ref{eq:r_max}, we consider only the QSO2s described as "non-compact" by \citet{fischer2018a}, as shown in Table \ref{tab:qso2s}. This yields ratios $R_{\rm [O~III]}/R^{\rm [O~III]}_{\rm max}\sim$0.51-1.11 for the non-compact QSO2s in \citetalias{trindadefalcao2021a}, with a mean value=0.85, and a standard deviation=0.21 (Fig. \ref{fig:ideal}, in red). If considering the compact targets, this ratio is $R_{\rm [O~III]}/R^{\rm [O~III]}_{\rm max}\sim$0.25-1.11 (mean value=0.71, and standard deviation=0.26, Fig. \ref{fig:ideal}, in green).

Given their comparable luminosities (see Tables \ref{tab:qso1s} and \ref{tab:qso2s}), we use the same $Q_{\rm [O~III]}$/$L_{\rm [O~III]}$ ratio used for the QSO2s to estimate $Q_{\rm [O~III]}$, $L^{\rm [O~III]}_{\rm bol}$, and $R^{\rm [O~III]}_{\rm max}$ for all PG QSO1s in this work (Table \ref{tab:qso1s_rmax}). The $R_{\rm [O~III]}/R^{\rm [O~III]}_{\rm max}$ ratios for the PG QSOs range from 0.50 to 2.93 (mean value=1.47, and standard deviation=0.90, Fig. \ref{fig:ideal}, in blue). While still reasonable, these ratios show a larger overall deviation from the "ideal" value of 1 compared to the QSO2s in \citetalias{trindadefalcao2021a} (Table \ref{tab:qso2s}, and Fig. \ref{fig:ideal}), which points to a larger scattering of the $R^{\rm [O~III]}_{\rm max}$ values for the QSO1s. 

\begin{table}
\begin{center}
\begin{tabular}{ccccc}
\multicolumn{1}{c}{}\\
\hline
\multicolumn{1}{l}{QSO1s}
&\multicolumn{1}{c}{$Q_{\rm [O~III]}$}
&\multicolumn{1}{c}{$L^{\rm [O~III]}_{\rm bol}$}
&\multicolumn{1}{c}{$R^{\rm [O~III]}_{\rm max}$}
&\multicolumn{1}{c}{\ul{$R_{\rm [O~III]}$}}
\\
& (photons~s$^{-1}$) & (erg~s$^{-1}$)& (kpc) & $R^{\rm [O~III]}_{\rm max}$ \\
\hline
\hline

   PG0026+129  & 2.5$\times10^{54}$ & $8.4\times10^{44}$ & 2.6 & 1.81 \\
   
  PG0052+251  & 5.3$\times10^{54}$& $1.8\times10^{45}$ & 3.9& 1.28\\
  
  PG0157+001  & 3.4$\times10^{54}$& $1.1\times10^{45}$ & 3.1 & 2.93\\
  
  PG0953+414  & 1.3$\times10^{55}$& $4.2\times10^{45}$ & 6.0 & 0.53\\
  
  PG1012+008 & 3.0$\times10^{54}$& $1.0\times10^{45}$ & 2.9 & 1.03\\
  
  PG1049\textminus005 & 1.6$\times10^{55}$& $5.6\times10^{45}$  & 6.8 & 0.50\\
  
  PG1307+085  & 4.8$\times10^{54}$& $1.6\times10^{45}$&3.7& 2.19 \\

 \hline
\end{tabular}
\end{center}
\caption{\textbf{Properties for the PG QSO1s from this work.} Note: (1) Galaxy name; (2) number of ionising photons per second emitted by the AGN; (3) Bolometric luminosity estimated from $L_{\rm [O~III]}$; (4) maximum NLR radius estimated using $L_{\rm [O~III]}$; (5) ratio between measured NLR size and estimated maximum NLR size.}
\label{tab:qso1s_rmax} 
\end{table}

Giving that $R_{\rm max}$ (Equation \ref{eq:r_max}) solely depends on the number of ionising photons \textit{Q}, the resulting scattering of $R^{\rm [O~III]}_{\rm max}$ for the QSO1s points to a similar trend in the values of $Q_{\rm [O~III]}$ for the sample, which, in turn, depends on $L_{\rm [O~III]}$ (Table \ref{tab:qso1s}). Below, we discuss possible explanations for this:

\begin{description}

    \item \textbf{\textit{i) [O~III] measurements affected by extinction?}} Since optical spectra covering both H$\alpha$ and H$\beta$ are not available for all PG QSO1s in the sample, we opted not to correct the [O~III] fluxes for extinction. However, in principle, using extinction-corrected [O~III] fluxes would yield larger $L_{\rm [O~III]}$, which could explain the differences between measured ($R_{\rm [O~III]}$) and predicted ($R^{\rm [O~III]}_{\rm max}$) NLR sizes. To check the magnitude of such effect, we examine existing SDSS spectra for 3 QSO1s in the sample: PG0157+001, PG1049-005, and PG1307+085. PG1049-005 and PG1307+085 have H$\alpha$/H$\beta$ $\approx$ 3.3. Assuming an intrinsic H$\alpha$/H$\beta$ ratio of 3.1 \citep[e.g.,][]{dong2008a}, and the extinction curve from \citet{cardelli1989a}, we obtain E(B-V) $\approx$0.057, which would increase the measured [O~III] fluxes in Table \ref{tab:qso1s} by a factor of $\sim$1.14 (following \citealt{seaton1979a}). For PG1057+001, the H$\alpha$ is contaminated by absorption from the atmospheric O$_{2}$ Fraunhofer A band. However, from the observed H$\beta$/H$\gamma\approx$ 2.47 we obtain E(B-V)$\approx$0.14, and an increase factor of $\sim$1.38 in the measured [O~III] fluxes listed in Table \ref{tab:qso1s}. The new (corrected for extinction) $R_{\rm [O~III]}/R^{\rm [O~III]}_{\rm max}$ values would range between 0.42-2.50 (for E(B-V)$\approx$0.14) for the QSO1s. Thus, uncorrected extinction would be insufficient to account for the differences in the measured and predicted NLR sizes.

    \smallskip
    \item \textbf{\textit{ii) NLR over-ionised \citep{baldwin1977a} in the PG QSO1s?}} Over-ionisation would reduce the amount of [O~III]-emitting gas in the NLR of these objects, yielding smaller $L_{\rm [O~III]}$ values, and therefore underestimated $Q_{\rm [O~III]}$. However, this possibility seems unlikely given the similar range in bolometric luminosity between the samples (see Table \ref{tab:qso1s_rmax} and \citetalias{trindadefalcao2021a}).

    \smallskip
    \item \textbf{\textit{iii) Differences in viewing angle (face-on \textit{vs.} edge on)?}} According to the Unified Model for AGNs \citep[e.g.][]{antonucci1985a}, the QSO1s would be observed "face-on", therefore showing "rounder" and smaller NLRs (given the expected foreshortening effects and viewing angle for type 1s, e.g., \citealt{mulchaey1996a}), as observed for PG0026+129, PG0953+414, PG1012+008, and PG1049-005 (Fig. \ref{fig:morphology1}). Meanwhile, the NLRs of type 2s would typically be observed as more "biconical", and more spatially-extended in the sky \citep[e.g.,][]{mulchaey1996a}. Projection effects arising from these differences would yield smaller NLR sizes for those targets seen "face-on", the PG QSO1s. However, as shown in Fig. \ref{fig:comparison}, the observed NLRs of these targets are \textit{larger} than those found for the QSO2s in \citetalias{trindadefalcao2021a}. 

    \smallskip
    \item \textbf{\textit{iv) Matter bounded \textit{vs.} radiation bounded?}} If the ionising bicone preferentially points into or out of the host galaxy plane in one of the samples, we should expect to observe either larger or smaller NLR sizes for such targets, respectively, depending on the amount of available [O~III]-emitting gas. However, given that the QSO2s in \citetalias{trindadefalcao2021a} were assumed to be pointing into the plane of the galaxy, but show smaller NLRs compared to the QSO1s in this work, this possibility seems unlikely. 

    \smallskip
    \item \textbf{\textit{v) \textbf{The overall spatial distribution of the [O~III]-emitting gas within the NLR is different between the samples.}}} Given that the bolometric correction factor applied to $L_{\rm [O~III]}$ in \citetalias{trindadefalcao2021a} was \textit{empirically} derived for type 2 objects \citep{lamastra2009a}, if the NLR gas in the QSO1s have overall different morphologies and distributions, then the same correction factor should not be used to estimate $Q_{\rm [O~III]}$ for the PG QSO1s. 
\end{description}

\begin{table*}
\begin{center}
\begin{tabular}{cccccc}
\multicolumn{1}{c}{}\\
\hline
\multicolumn{1}{l}{QSO1s}
&\multicolumn{1}{c}{$L_{\rm {5100\AA}}$}
&\multicolumn{1}{c}{$Q_{5100}$}

&\multicolumn{1}{c}{$L^{\rm 5100\AA}_{\rm {bol}}$}
&\multicolumn{1}{c}{$R^{\rm 5100\AA}_{\rm max}$}
&\multicolumn{1}{c}{\ul{$R_{\rm [O~III]}$}}
\\
& (erg~s$^{-1}$) & (photons~s$^{-1}$)& (erg~s$^{-1}$)& (kpc) & $R^{\rm 5100\AA}_{\rm max}$ \\
\hline
\hline
  PG0026+129 & $5.4\times10^{44}$ &1.9$\times10^{55}$ & $6.2\times10^{45}$  & 7.2 & 0.63 \\
  PG0052+251 & $2.5\times10^{44}$ & 9.9$\times10^{54}$&$3.3\times10^{45}$  & 5.2 & 1.06 \\
  PG0157+001 & $2.1\times10^{44}$ & 8.7$\times10^{54}$&$2.9\times10^{45}$  & 4.9 & 1.86 \\
  PG0953+414 & $1.2\times10^{45}$ & 3.6$\times10^{55}$&$1.2\times10^{46}$ & 10.0 & 0.35 \\
  PG1012+008 & $3.6\times10^{44}$ & 1.3$\times10^{55}$&$4.4\times10^{45}$ & 6.1 & 0.49\\
  PG1049\textminus005 & $1.0\times10^{45}$ & 3.0$\times10^{55}$& $1.0\times10^{46}$  & 9.1 & 0.49 \\
  PG1307+085 & $3.4\times10^{44}$ & 1.3$\times10^{55}$& $4.2\times10^{45}$ & 5.9 & 1.37 \\
\hline
\end{tabular}
\end{center}
\caption{\textbf{Continuum (5100\AA) measurements for the PG QSO1s.}}
\label{tab:qso1s_lbol_cont} 
\end{table*}

In summary, using photoionisation modeling and Equation \ref{eq:r_max} to estimate the maximum NLR sizes in the QSO1s and QSO2s, we find size estimates that agree well with direct measurements for the QSO2s from \citetalias{trindadefalcao2021a}, but are generally underestimated for the current sample of QSO1s. Such an underestimation may arise from different spatial distributions of the [O~III]-emitting gas within the NLR of these samples, which in turn would affect estimates of the bolometric luminosity and number of ionising photons in these sources.

\subsection{Bolometric Luminosity}
\label{sec:bolometric_luminosity}

In the previous section, we examine the possibility that the NLR gas in the QSO1s may show different spatial distributions than in the QSO2s from \citetalias{trindadefalcao2021a}. A direct consequence of this is that estimating the "true" AGN ionising power ($Q_{\rm [O~III]}$ and $L^{\rm [OIII]}_{\rm bol}$) from $L_{\rm [OIII]}$ and a bolometric correction factor for the QSO1s may yield inaccurate results. Fortunately, for type 1 objects, the AGN ionising power can be measured directly from the radiation arising from the nucleus. In this case, the AGN's continuum luminosity can be used to estimate $Q$, and $L{\rm bol}$. 

In this study, we adopt the formalism from \citet{netzer2019a} to derive a bolometric correction factor for the QSO1s in our sample. This correction factor ($k_{\rm bol}$) was developed combining observed X-ray AGN properties and theoretical calculations of optically thick, geometrically thin accretion disks, and is valid over a wide range of black hole mass, spin, and accretion rate. For the QSO1s, we use the simple power-law approximation of $k_{\rm bol}$ for the observed 5100\AA~continuum luminosity, $L_{\rm {5100\AA}}$, from \citet{netzer2019a}: 

\begin{equation}
\label{eq:kbol}
    k_{\rm bol} = 40 \times \left(\frac{L_{\rm {5100\AA}}}{10^{42} \, \text{erg/sec}}\right)^{-0.2}
\end{equation}

This method has the advantage of relying solely on accretion disk properties, and it does not depend on specific properties of the NLR gas. 

The 5100\AA~continuum luminosities for the QSO1s are measured as discussed in Section \ref{sec:data_reduction}, and the new derived bolometric luminosities from $L_{\rm {5100\AA}}$, $L^{\rm {5100\AA}}_{\rm bol}$ are listed in Table \ref{tab:qso1s_lbol_cont}. The values of $L^{\rm {5100\AA}}_{\rm bol}$ in Table \ref{tab:qso1s_lbol_cont} are consistently larger than those derived from [O~III] ($L^{\rm [O~III]}_{\rm bol}$, Table \ref{tab:qso1s_rmax}), which supports the idea of different spatial distributions for the [O~III]-emitting gas between the QSO1s and QSO2s samples. 

We now recalculate $R_{\rm max}$ (Equation \ref{eq:r_max}) for the QSO1s using the bolometric luminosity values derived from the continuum luminosity at 5100\AA, $L^{\rm {5100\AA}}_{\rm bol}$ (Table \ref{tab:qso1s_lbol_cont}). We also compute new $R_{\rm [O~III]}/R^{\rm 5100\AA}_{\rm max}$ values, which range from 0.35 to 1.86 (mean value=0.89, standard deviation=0.56, Fig. \ref{fig:ideal}, in orange), and show smaller deviation than those derived using $L^{\rm [O~III]}_{\rm bol}$ (Table \ref{tab:qso1s_rmax}, Fig. \ref{fig:ideal}, in blue). The lowest $R_{\rm [O~III]}/R^{\rm 5100\AA}_{\rm max}$ ratios are observed for PG0026+29 (0.63), PG0953+414 (0.35), PG1012+008 (0.49), and PG1049-005 (0.49). These are the four AGNs with more symmetric NLRs (see Table \ref{tab:qso1s}), and in which we expect to see more significant projection effects (i.e., their deprojected NLR sizes are expected to be larger than measured in Table \ref{tab:qso1s}).

Comparing the different $R_{\rm [O~III]}/R^{\rm  [O~III]}_{\rm max}$ trends in Fig. \ref{fig:ideal} (blue for the QSO1s and red for the non-compact QSO2s), the different curves reveal differences between the samples. Specifically, using $R^{\rm  [O~III]}_{\rm max}$ to estimate the NLR sizes for the samples results in significantly larger sizes for the QSO1s. However, when estimating the bolometric luminosity from the 5100\AA~continuum luminosity for the QSO1s ($R_{\rm [O~III]}/R^{\rm 5100\AA}_{\rm max}$, in orange) yields a distribution that more closely aligns with that of the QSO2s.


\subsection{Ionised Gas Masses}
\label{sec:Ionised Gas Masses}
We examine the morphology and spatial distribution of the NLR gas in the QSO1s by means of [O~III] mass distribution profiles. We first use the continuum-subtracted [O~III] fluxes (see Section \ref{sec:data_reduction}) to construct [O~III] flux radial profiles for each QSO1. We then compute the masses of [O~III] gas within each annular region, by assuming a density radial profile. As discussed in \citetalias{trindadefalcao2021a} and \citet{revalski2022a}, gas densities in the NLR can be estimated via spatially-resolved images or spectra, combined with single-component photoionisation models. 

In \citetalias{trindadefalcao2021a}, we used a single-component model with log($U$)=-2 to model the [O~III] gas in the QSO2s. However, for the QSO1s in this work we assume that the NLR extends up to the distance where gas with log($n_{\rm H}$)=2 is characterised by log($U$)$<$-3 (following Equation \ref{eq:r_max}). Therefore, for consistency, we build photoionisation models assuming log($U$)=-2 up to where the NLR gas has log($n_{\rm H}$)=2, after which we keep log($n_{\rm H}$)=2 while the ionisation parameter is decreased to log($U$)=-3. Our photoionisation models use \texttt{CLOUDY} \citep{ferland2017a}, and consider the same SED and abundances as in \citetalias{trindadefalcao2021a} (see Section \ref{sec:nlr_sizes_t1_t2}). We assume that the emission-line gas is radiation-bounded, with model integration stopping when the electron temperature drops below 4000K. As the temperature drops below 4000K the gas becomes molecular, hence there is very little atomic line emission.

Following \citetalias{trindadefalcao2021a}, we obtain the emitting area of the emission-line gas at each annulus position, 
\begin{equation}
\label{eq:area}
    A(r)=\frac{L_{\rm [O~III]}}{F_{\rm [O~III]}(r)}
\end{equation}

\noindent where $L_{\rm [O~III]}$ is the observed [O~III] luminosity in erg s$^{-1}$, and $F_{\rm [O~III]}(r)$ is the emitted flux estimated by \texttt{CLOUDY} at a given radius $r$. 

The column densities, $N_{\rm H}$ (cm$^{-2}$), computed by the models are then used to calculate the mass of emitting gas at each radius $r$:

\begin{equation}
\label{eq:mass}
M(r) = N_{\rm H}~1.4 \textit{m}_{\rm H}~\left(\frac{L_{\rm [O~III]}}{F_{\rm [O~III]}(r)}\right)
\end{equation}

\noindent
where $m_{\rm H}$ is the atomic mass of hydrogen and the factor of 1.4 accounts for the contribution from helium.  

The uncertainties inherent to this method are discussed in detail in \citetalias{trindadefalcao2021a}. Among them are the assumptions that the SEDs and elemental abundances of the two samples are the same as those used for Mrk~34 in \citetalias{trindadefalcao2021a}. In addition to the uncertainties in our models, there are inherent uncertainties due to resolution limits and projection effects. Nevertheless, the resulting mass profiles allow us to probe differences among the PG QSO1s in the sample.

\begin{figure*}
    \centering
    \includegraphics[width=\textwidth]{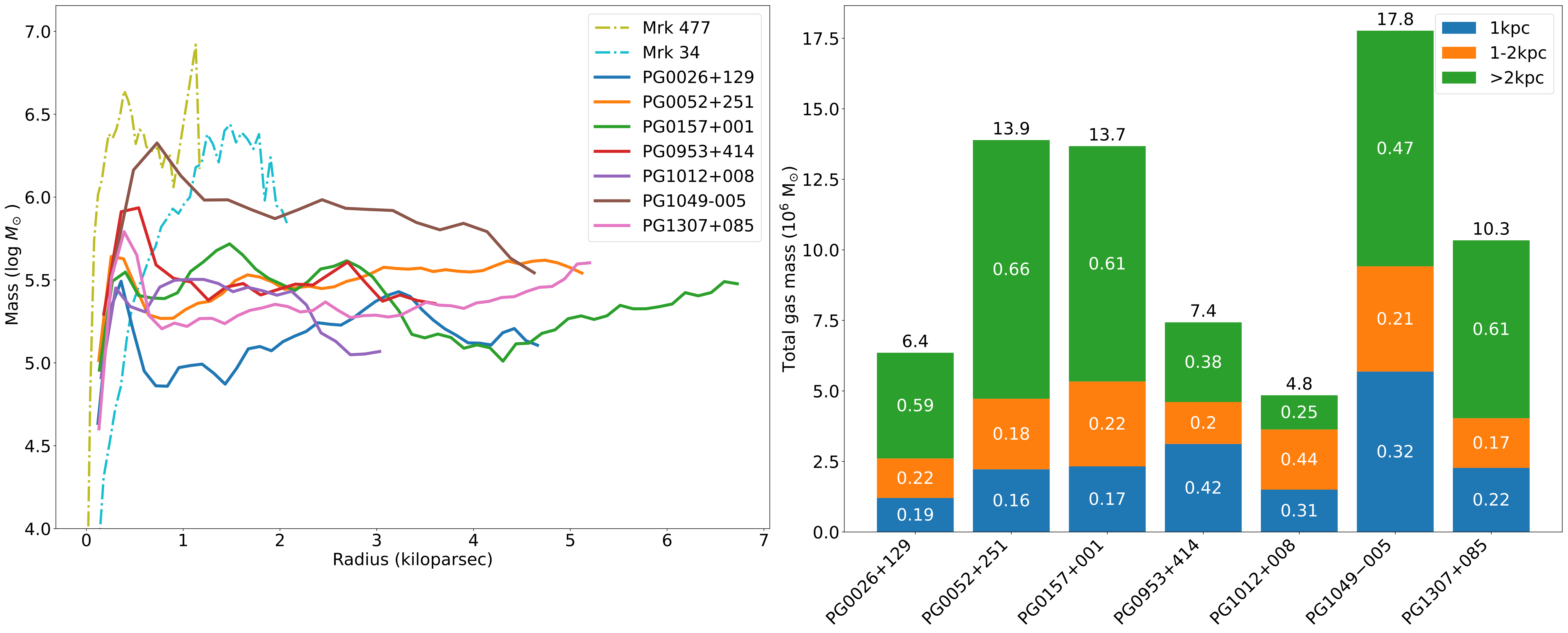}
    \caption{\textbf{Left:} Spatially-resolved [O~III] gas mass profiles for the PG QSO1s in our sample. The masses are the masses in each radial bin. We also show the mass profiles of 2 QSO2s from \citetalias{trindadefalcao2021a}. \textbf{Right:} Histogram of the spatial distribution and total ionised gas masses for the PG QSO1s in our sample. The values within the bars are the fraction of total mass from 0-1 kpc (blue), 1-2 kpc (orange) and $> $ kpc (green). The values above the bars are the total mass in unit of 10$^{6}$ $M_\odot$.}
    \label{fig:masses}
\end{figure*}

The resulting mass profiles are shown in the left panel of Fig. \ref{fig:masses}. All seven PG QSO1s show significant spatially-extended mass profiles ($\geq$2 kpc). Such extended mass profiles are similar (i.e., extending to distances $>$2 kpc) to those derived for Mrk 34 (dashed cyan line in Fig. \ref{fig:masses}) and Mrk 78 in \citet{revalski2021a}, the two highest luminosity AGNs in their sample. 

Compared to the QSO2s in \citetalias{trindadefalcao2021a}, the PG QSO1s show even more extended NLRs than that of Mrk 34 ($\sim$2 kpc, dashed cyan line in Fig. \ref{fig:masses}), and Mrk 477 (dashed olive line), even for PG1012+008, which has the smallest NLR in our sample (3.0 kpc, Table \ref{tab:qso1s}). However, when considering their integrated gas mass, the QSO2s show larger ionised gas masses than the QSO1s. For example, the total mass of ionised gas in the NLR of Mrk 34 (within 2 kpc) is $\sim3.4\times10^{7}~M_{\odot}$, which is much larger than the gas masses obtained for any of the PG QSO1s in our sample (Fig. \ref{fig:masses}). 

Fig. \ref{fig:masses} (right panel) shows a mass histogram of the results presented in the left panel for the QSO1s, including the total mass of ionised gas, as well as the percentage of gas mass at different radii from the AGN, $r<1$ kpc (blue), $1<r\leq2$ kpc (orange), and $r>$2 kpc (green). PG1049-005 has the largest mass of ionised gas, $M\sim1.8\times10^{7}M_{\odot}$, with the highest fraction ($\sim47\%$) located at distances $r>$2 kpc. Four of the other PG QSO1s, PG0026+129, PG0052+251, PG0157+001, and PG1307+085 show similar mass distribution (see Table \ref{tab:masses}).  

\subsection{Type 1 \textit{vs.} Type 2 Dichotomy}
\label{sec:dichotomy}
Our results in Sections \ref{sec:results} and \ref{sec:discussion} suggest differences between the QSO1s analysed in this work, and the QSO2s in \citetalias{trindadefalcao2021a}: 

\smallskip 
\textbf{(1)} The type 1 objects in Fig. \ref{fig:comparison} follow a steeper $R_{\rm [O~III]}$-$L_{\rm[O~III]}$ relation than that of the type 2s analysed, $\gamma_{1}$=0.57$\pm0.05$, and $\gamma_{2}$=0.48$\pm0.02$ (see Equations \ref{eq:trindade_relation1} and \ref{eq:trindade_relation2}). This suggests larger NLRs for the type 1s than the type 2s considered in Section \ref{sec:relation_r_and_L}. This is also evident when comparing the measured NLR sizes for the PG QSO1s (Table \ref{tab:qso1s}) with those for the QSO2s in \citetalias{trindadefalcao2021a} (Table \ref{tab:qso2s}); 

\smallskip 
\textbf{(2)} When calculating $R_{\rm max}$ in Section \ref{sec:Constraints_on_NLR_Sizes}, we estimate $Q_{\rm [O~III]}$ for the QSO2s from $L_{\rm[O~III]}$, which returns an excellent agreement with the measured NLR sizes (i.e., a $R_{\rm [O~III]}/R^{\rm [O~III]}_{\rm max}$ ratio close to 1, see Fig. \ref{fig:ideal}), $R_{\rm [O~III]}/R^{\rm [O~III]}_{\rm max}$=0.51-1.11. However, the same approach applied to the PG QSO1s returns more scattered values for NLR sizes (Fig. \ref{fig:ideal}, in blue), $R_{\rm [O~III]}/R^{\rm [O~III]}_{\rm max}$=0.50-2.93. A better agreement (i.e., closer to the ideal value of 1) is obtained for the QSO1s when $L_{\rm bol}$ (and $Q$) is derived from the 5100\AA~continuum luminosity instead ($R_{\rm [O~III]}/R^{\rm 5100\AA}_{\rm max}$=0.35-1.86, Fig. \ref{fig:ideal}, in orange). These suggest differences in the spatial distribution of the NLR gas between the two samples;  

\smallskip 
\textbf{(3)} The mass profiles for the PG QSO1s (Fig. \ref{fig:masses}, left) reveal significantly extended radial profiles for the PG QSO1s, with most of their mass concentrated at larger radii in five of seven AGNs (i.e., not for PG 0953+414, and PG 1012+008). When compared to the QSO2s in \citetalias{trindadefalcao2021a}, the PG QSO1s show larger NLR sizes (by a factor of $\sim$3, see Tables \ref{tab:qso1s} and \ref{tab:qso2s}), but smaller gas mass (see Fig. \ref{fig:masses}, and Tables \ref{tab:qso2s}, \ref{tab:masses}). The lower gas mass may arise from underestimating $L_{\rm bol}$ for the PG QSO1s, when applying the bolometric correction from \citet{lamastra2009a} to $L_{\rm[O~III]}$ (see Section \ref{sec:nlr_sizes_t1_t2}).

Assuming a direct proportionality between bolometric luminosity ($L_{\rm bol}$) and the ionising luminosity (\textit{Q}), the underestimation of $L^{\rm[O~III]}_{\rm bol}$ suggests less [O~III] gas in the NLR of the QSO1s. To test this, we use $L^{\rm5100\AA}_{\rm {bol}}$ (Table \ref{tab:qso1s_lbol_cont}) to estimate the "true" $L_{\rm[O~III]}$ for the PG QSOs, by scaling it by the \citet{lamastra2009a} correction. This effectively calculates the $L_{\rm[O~III]}$ that QSO1s would have if their NLR showed the same [O~III] distribution as the QSO2s (see Fig. \ref{fig:comparison}). The resulting slope of $\gamma_{1}$=0.49$\pm0.05$, aligns with both the QSO2s slope and the overall AGN slope in Fig. \ref{fig:comparison}.

As discussed in Section \ref{sec:Constraints_on_NLR_Sizes}, \citet{fischer2018a} identify two different groups of QSO2s within the full sample in \citetalias{trindadefalcao2021a}. They suggest that the differences found among the QSO2s are a consequence of different concentrations of gas in the inner region around the AGNs, which would affect the degree of attenuation of the ionising radiation to the NLR \citep{fischer2018a}. In this scenario, strong outflows (if present) could drive the most of the inner gas away from the inner regions \citep{das2005a}, reducing the amount of emission-line gas at small radii. \citet{fischer2018a} suggest that this could explain the observed transition seen in the sample, from targets with spatially compact NLRs, such as Mrk 477 ($\sim$1 kpc, Fig. \ref{fig:masses} and \citetalias{trindadefalcao2021a}), to those with more spatially-extended NLRs, such as Mrk 34 ($\sim$2 kpc, Fig. \ref{fig:masses} and \citetalias{trindadefalcao2021a}).

With our results for the PG QSO1s, we now propose a complementary interpretation to that of \citet{fischer2018a}. The differences observed here between QSO1s and QSO2s suggest that we may be observing these objects at different times in their evolution. Given that QSO1s seem to have more extended, albeit less massive NLRs, it is possible these AGNs have already passed their "blown-out" stage, in which gas concentrated in the inner regions of the AGN has now been ionised and driven away to larger radii. In this scenario, AGNs with more compact NLR morphologies, such as the QSO2 Mrk 477 \citep{fischer2018a}, will eventually turn into AGNs with more extended NLRs, such as the QSO2 Mrk 34 \citep{fischer2018a} and the QSO1 PG0026+129, and ultimately into AGNs with very extended, but less massive NLRs, as seen for PG0157+001.

\begin{table*}
\begin{center}
\caption{\textbf{[O~III] gas masses for the PG QSOs in our sample.}}
\label{tab:masses} 

\begin{tabular}{lcccc}
\multicolumn{1}{c}{}\\
\hline
\multicolumn{1}{l}{QSO1s}
&\multicolumn{1}{c}{Total Ionised Gas Mass}
&\multicolumn{1}{c}{Enclosed Mass Fraction}
&\multicolumn{1}{c}{Enclosed Mass Fraction}
&\multicolumn{1}{c}{Enclosed Mass Fraction}

\\
&($10^{6}M_{\odot}$)&($r\leq$1 kpc) & ($1<r\leq2$ kpc)& ($r>$2 kpc) \\
\hline
\hline
  PG0026+129  & 5.9 & 19\% & 22\%& \textbf{59\%}\\
  PG0052+251 &12.9&16\%&18\%&\textbf{66\%}\\
  PG0157+001 & 12.7&17\%&22\%&\textbf{61\%}\\
  PG0953+414 & 6.9&\textbf{42\%}&20\%&38\%\\
  PG1012+008 &4.5&31\%&\textbf{44\%}&25\%\\
  PG1049\textminus005 &16.5&32\%&21\%&\textbf{47\%}\\
  PG1307+085 &9.6&22\%&17\%&\textbf{61\%}\\
\hline
\end{tabular}
\end{center}
\caption*{Total [O~III] gas masses and enclosed mass fractions for the PG QSOs in our sample. We show the bulk of gas mass in boldface for each object.}

\end{table*}

\section{Conclusions}
\label{sec:conclusions}

We present \textit{HST} [O~III] 5007\AA~narrow-band images of the seven brightest PG QSO1s from the sample of Bright QSO Survey \citep[][]{schmidt1983a, kellermann1989a}. Our main findings are summarized below:

\begin{description}
    \item \textbf{1)} We detect spatially-extended NLRs in all seven sources, from 3-9 kpc. The largest NLRs are observed for PG1307+085 ($R_{\rm [O~III]}$=8.1 kpc), and PG0157+001 ($R_{\rm [O~III]}$=9.1 kpc). Regarding their morphology, PG QSO1s showing more compact NLRs ($<$5 kpc) also show more round gas distribution around the central nucleus, while those showing more extended NLRs ($>$5 kpc) have more asymmetric and elongated gas distributions. 

    \smallskip
    \item \textbf{2)} Overall, when combined with results for lower-redshift and lower-luminosity AGNs, we find that the size of the NLRs scale with luminosity in these objects. We find that the size of the NLR increases with $L_{\rm [O~III]}$ as $\rm log (R_{\rm [O~III]})_{all} = (0.51\pm0.03)log(L_{\rm [O~III]}) -18.13\pm1.06$, over the range 39.5 $<$ log $L_{\rm [O~III]}<$ 43.4 erg ${\rm s^{-1}}$. However, when analyzed separately, type 1s show a much steeper dependence ($\rm log (R_{\rm [O~III]})_{1} = (0.57\pm 0.05)log(L_{\rm [O~III]}) -20.86\pm2.35$) of NLR extent on [O~III] luminosity than type 2 AGNs ($\rm log (R_{\rm [O~III]})_{2} = (0.48\pm0.02)log(L_{\rm [O~III]}) -17.13\pm1.07$, see also Fig. \ref{fig:comparison}), suggesting larger NLRs, but somehow lower gas masses.

    \smallskip
    \item \textbf{3)} We use \texttt{CLOUDY} photoionisation models to obtain upper limits on NLR sizes. We constrain the ionisation parameter (log\textit{U}=-3), and density ($n_{\rm H}\sim$ 100 cm$^{-3}$) of the gas, and use an ionising luminosity derived from $L_{\rm [O~III]}$ \citep{lamastra2009a}. Compared to measured NLR sizes, our estimates agree well for the QSO2s ($R_{\rm [O~III]}/R^{\rm [O~III]}_{\rm max}$=0.51-1.11, Fig. \ref{fig:ideal}), but overestimate those in the PG QSO1s ($R_{\rm [O~III]}/R^{\rm [O~III]}_{\rm max}$=0.50-2.93, Fig. \ref{fig:ideal}). We find a better agreement when estimating $R_{\rm max}$ from an ionising luminosity (and bolometric luminosity) derived from the 5100\AA~continuum luminosity instead ($R_{\rm [O~III]}/R^{\rm 5100\AA}_{\rm max}$=0.35-1.86, Fig. \ref{fig:ideal}), pointing to an underestimation of $L_{\rm [O~III]}$ for the QSO1s. 
    
    \smallskip
    \item \textbf{4)} We calculate spatially-resolved mass profiles for all seven QSO1s. These show significantly spatially-extended NLRs in all objects, with five of the targets showing more mass at distances $>$2 kpc. We derive integrated gas masses in the range 4.5$\times10^{6}M_{\odot}$ (PG1012+008) - $\sim$16.5$\times10^{6}M_{\odot}$ (PG1049-005). Note that the total ionised gas mass for the QSO1s are smaller than those found for the QSO2s Mrk~34 and Mrk~477 in \citetalias{trindadefalcao2021a}.

    \smallskip
    \item \textbf{5)} Given our results and previous discussion, we propose an evolutionary scenario in which compact-morphology AGNs, such as Mrk 477 \citep{fischer2018a}, may evolve into AGNs with more spatially-extended NLRs, such as Mrk 34, and ultimately into AGNs with more extended NLRs, albeit with less gas mass, like the PG QSO1s analyzed here.

\end{description}

We will analyze the \textit{HST}/STIS long-slit spectroscopic data for the PG QSO1s in a forthcoming paper (Trindade Falcao et al., \textit{in preparation}). Our analysis will yield detailed information on the kinematics of the ionised gas, including FWHM, and outflow velocities. We will also analyze any potential correlations between the NLR morphology of these AGNs and their evolutionary path, specifically focusing on the presence or absence of high [O~III] FWHM in the vicinity of the nucleus \citep{fischer2018a}. We will compare our results to those reported in \citet{fischer2018a}.

\section*{Acknowledgements}

Support for this work was provided by NASA through grant number HST-GO-13728.001-A from the Space Telescope Science Institute, which is operated by AURA, Inc., under NASA contract NAS 5-26555. Basic research at the Naval Research Laboratory is funded by 6.1 base funding. T.C.F. was supported by an appointment to the NASA Postdoctoral Program at the NASA Goddard Space Flight Center, administered by the Universities Space Research Association under contract with NASA. T.S.-B. acknowledges support from the Brazilian institutions CNPq (Conselho Nacional de Desenvolvimento Científico e Tecnológico) and FAPERGS (Fundação de Amparo à Pesquisa do Estado do Rio Grande do Sul).  L.C.H. was supported by the National Science Foundation of China (11721303, 11991052, 12011540375, 12233001), the National Key R\&D Program of China (2022YFF0503401), and the China Manned Space Project (CMS-CSST-2021-A04, CMS-CSST-2021-A06). M.V. gratefully acknowledges financial support from the Danish Council for Independent Research via grant no. DFF 4002-00275, 8021-00130 and 3103-00146. M.W. acknowledges support of a Leverhulme Emeritus Fellowship, EM-2021-064.\par 
This research has made use of the NASA/IPAC Extragalactic Database (NED), which is operated by the Jet Propulsion Laboratory, California Institute of Technology, under contract with the National Aeronautics and Space Administration.

\section*{Data Availability}

Based on observations made with the NASA/ESA Hubble Space Telescope, and available from the Hubble Legacy Archive, which is a collaboration between the Space Telescope Science Institute (STScI/NASA), the Space Telescope European Coordinating Facility (ST-ECF/ESAC/ESA) and the Canadian Astronomy Data Centre (CADC/NRC/CSA).



\bibliographystyle{mnras}
\bibliography{anna_bibliography} 

\begin{thebibliography}{}
\makeatletter
\relax
\def\mn@urlcharsother{\let\do\@makeother \do\$\do\&\do\#\do\^\do\_\do\%\do\~}
\def\mn@doi{\begingroup\mn@urlcharsother \@ifnextchar [ {\mn@doi@}
  {\mn@doi@[]}}
\def\mn@doi@[#1]#2{\def\@tempa{#1}\ifx\@tempa\@empty \href
  {http://dx.doi.org/#2} {doi:#2}\else \href {http://dx.doi.org/#2} {#1}\fi
  \endgroup}
\def\mn@eprint#1#2{\mn@eprint@#1:#2::\@nil}
\def\mn@eprint@arXiv#1{\href {http://arxiv.org/abs/#1} {{\tt arXiv:#1}}}
\def\mn@eprint@dblp#1{\href {http://dblp.uni-trier.de/rec/bibtex/#1.xml}
  {dblp:#1}}
\def\mn@eprint@#1:#2:#3:#4\@nil{\def\@tempa {#1}\def\@tempb {#2}\def\@tempc
  {#3}\ifx \@tempc \@empty \let \@tempc \@tempb \let \@tempb \@tempa \fi \ifx
  \@tempb \@empty \def\@tempb {arXiv}\fi \@ifundefined
  {mn@eprint@\@tempb}{\@tempb:\@tempc}{\expandafter \expandafter \csname
  mn@eprint@\@tempb\endcsname \expandafter{\@tempc}}}

\bibitem[\protect\citeauthoryear{{Antonucci} \& {Miller}}{{Antonucci} \&
  {Miller}}{1985}]{antonucci1985a}
{Antonucci} R.,  {Miller} J.~S.,  1985, The Astrophysical Journal, 297, 621

\bibitem[\protect\citeauthoryear{{Bae} \& {Woo}}{{Bae} \&
  {Woo}}{2016}]{bae2016a}
{Bae} H.-J.,  {Woo} J.-H.,  2016, The Astrophysical Journal, 828, 97

\bibitem[\protect\citeauthoryear{{Bahcall}, {Kirhakos}, {Saxe}  \&
  {Schneider}}{{Bahcall} et~al.}{1997}]{bahcall1997a}
{Bahcall} J.~N.,  {Kirhakos} S.,  {Saxe} D.~H.,   {Schneider} D.~P.,  1997, The
  Astrophysical Journal, 479, 642

\bibitem[\protect\citeauthoryear{{Baldwin}}{{Baldwin}}{1977}]{baldwin1977a}
{Baldwin} J.~A.,  1977, ApJ, 214

\bibitem[\protect\citeauthoryear{Behroozi, Wechsler, Hearin  \&
  Conroy}{Behroozi et~al.}{2019}]{behroozi2019a}
Behroozi P.,  Wechsler R.~H.,  Hearin A.~P.,   Conroy C.,  2019, MNRAS, 488,
  3143

\bibitem[\protect\citeauthoryear{{Bennert}, {Falcke}, {Schulz}, {Wilson}  \&
  {Wills}}{{Bennert} et~al.}{2002}]{bennert2002a}
{Bennert} N.,  {Falcke} H.,  {Schulz} H.,  {Wilson} A.~S.,   {Wills} B.~J.,
  2002, The Astrophysical Journal, 574, L105

\bibitem[\protect\citeauthoryear{{Boroson} \& {Green}}{{Boroson} \&
  {Green}}{1992}]{boroson1992a}
{Boroson} T.~A.,  {Green} R.~F.,  1992, The Astrophysical Journal Supplement,
  80, 109

\bibitem[\protect\citeauthoryear{{Cardelli}, {Clayton}  \& {Mathis}}{{Cardelli}
  et~al.}{1989}]{cardelli1989a}
{Cardelli} J.~A.,  {Clayton} G.~C.,   {Mathis} J.~S.,  1989, The Astrophysical
  Journal, 345, 245

\bibitem[\protect\citeauthoryear{{Crenshaw} \& {Kraemer}}{{Crenshaw} \&
  {Kraemer}}{2005}]{crenshaw2005a}
{Crenshaw} D.~M.,  {Kraemer} S.~B.,  2005, The Astrophysical Journal, 625, 680

\bibitem[\protect\citeauthoryear{{Crenshaw}, {Kraemer}  \& {George}}{{Crenshaw}
  et~al.}{2003}]{crenshaw2003a}
{Crenshaw} D.~M.,  {Kraemer} S.~B.,   {George} I.~M.,  2003, Annual Review of
  Astronomy and Astrophysics, 41, 117

\bibitem[\protect\citeauthoryear{{Crenshaw}, {Kraemer}, {Schmitt}, {Jaff{\'e}},
  {Deo}, {Collins}  \& {Fischer}}{{Crenshaw} et~al.}{2010}]{crenshaw2010a}
{Crenshaw} D.~M.,  {Kraemer} S.~B.,  {Schmitt} H.~R.,  {Jaff{\'e}} Y.~L.,
  {Deo} R.~P.,  {Collins} N.~R.,   {Fischer} T.~C.,  2010, The Astronomical
  Journal, 139, 871

\bibitem[\protect\citeauthoryear{{Das} et~al.,}{{Das} et~al.}{2005}]{das2005a}
{Das} V.,  et~al., 2005, The Astronomical Journal, 130, 945

\bibitem[\protect\citeauthoryear{{Di Matteo}, {Springel}  \& {Hernquist}}{{Di
  Matteo} et~al.}{2005}]{dimatteo2005a}
{Di Matteo} T.,  {Springel} V.,   {Hernquist} L.,  2005, Nature, 433, 604

\bibitem[\protect\citeauthoryear{{Dong}, {Wang}, {Wang}, {Yuan}, {Zhou}, {Dai}
  \& {Zhang}}{{Dong} et~al.}{2008}]{dong2008a}
{Dong} X.,  {Wang} T.,  {Wang} J.,  {Yuan} W.,  {Zhou} H.,  {Dai} H.,   {Zhang}
  K.,  2008, Monthly Notices of the Royal Astronomical Society, 383, 581

\bibitem[\protect\citeauthoryear{{Fabian}}{{Fabian}}{2012}]{fabian2012a}
{Fabian} A.~C.,  2012, Annual Review of Astronomy and Astrophysics, 50, 455

\bibitem[\protect\citeauthoryear{{Falcke}, {Wilson}  \& {Simpson}}{{Falcke}
  et~al.}{1998}]{falcke1998a}
{Falcke} H.,  {Wilson} A.~S.,   {Simpson} C.,  1998, ApJ, 502, 199

\bibitem[\protect\citeauthoryear{{Ferland} et~al.,}{{Ferland}
  et~al.}{2017}]{ferland2017a}
{Ferland} G.~J.,  et~al., 2017, Revista Mexicana de Astronomia y Astrofisica,
  49, 1379

\bibitem[\protect\citeauthoryear{{Fischer}, {Crenshaw}, { Kraemer}  \&
  {Schmitt}}{{Fischer} et~al.}{2013}]{fischer2013a}
{Fischer} T.~C.,  {Crenshaw} D.~M.,  { Kraemer} S.~B.,   {Schmitt} H.~R.,
  2013, The Astrophysical Journal, 209, 1

\bibitem[\protect\citeauthoryear{{Fischer} et~al.,}{{Fischer}
  et~al.}{2017}]{fischer2017a}
{Fischer} T.~C.,  et~al., 2017, The Astrophysical Journal, 834, 30

\bibitem[\protect\citeauthoryear{{Fischer} et~al.,}{{Fischer}
  et~al.}{2018}]{fischer2018a}
{Fischer} T.~C.,  et~al., 2018, The Astrophysical Journal, 856, 102

\bibitem[\protect\citeauthoryear{Fischer et~al.,}{Fischer
  et~al.}{2021}]{fischer2021a}
Fischer T.~C.,  et~al., 2021, \mn@doi [The Astrophysical Journal]
  {10.3847/1538-4357/abca3c}, 906, 88

\bibitem[\protect\citeauthoryear{{Gandhi} et~al.,}{{Gandhi}
  et~al.}{2014}]{gandhi2014a}
{Gandhi} P.,  et~al., 2014, The Astrophysical Journal, 792, 117

\bibitem[\protect\citeauthoryear{{Greene}, {Zakamska}, {Ho}  \&
  {Barth}}{{Greene} et~al.}{2011}]{greene2011a}
{Greene} J.~E.,  {Zakamska} N.~L.,  {Ho} L.~C.,   {Barth} A.~J.,  2011, The
  Astrophysical Journal, 732, 9

\bibitem[\protect\citeauthoryear{G{\"u}ltekin et~al.,}{G{\"u}ltekin
  et~al.}{2009}]{gultekin2009a}
G{\"u}ltekin K.,  et~al., 2009, \mn@doi [The Astrophysical Journal]
  {10.1088/0004-637X/698/1/198}, 698, 198

\bibitem[\protect\citeauthoryear{{Hopkins} \& {Elvis}}{{Hopkins} \&
  {Elvis}}{2010}]{hopkins2010a}
{Hopkins} P.~F.,  {Elvis} M.,  2010, Monthly Notices of the Royal Astronomical
  Society, 401, 7

\bibitem[\protect\citeauthoryear{{Kellermann}, {Sramek}, {Schmidt}, {Shaffer}
  \& {Green}}{{Kellermann} et~al.}{1989}]{kellermann1989a}
{Kellermann} K.~I.,  {Sramek} R.,  {Schmidt} M.,  {Shaffer} D.~B.,   {Green}
  R.,  1989, The Astrophysical Journal, 98, 1195

\bibitem[\protect\citeauthoryear{{King} \& {Pounds}}{{King} \&
  {Pounds}}{2015}]{king2015a}
{King} A.,  {Pounds} K.,  2015, Annual Review of Astronomy and Astrophysics,
  53, 115

\bibitem[\protect\citeauthoryear{{Koraktar}}{{Koraktar}}{1999}]{koratkar1999a}
{Koraktar} A.,  1999, Astronomical Society of the Pacific, 111, 1

\bibitem[\protect\citeauthoryear{{Kraemer} et~al.,}{{Kraemer}
  et~al.}{2012}]{kraemer2012a}
{Kraemer} S.~B.,  et~al., 2012, The Astrophysical Journal, 751, 84

\bibitem[\protect\citeauthoryear{{Lamastra}, {Bianchi}, {Matt}, {Perola},
  {Barcons}  \& {Carrera}}{{Lamastra} et~al.}{2009}]{lamastra2009a}
{Lamastra} A.,  {Bianchi} S.,  {Matt} G.,  {Perola} G.~C.,  {Barcons} X.,
  {Carrera} F.~J.,  2009, Astronomy and Astrophysics, 504, 73

\bibitem[\protect\citeauthoryear{{Leipski}, {Falcke}, {Bennert}  \&
  {H{\"u}ttemeister}}{{Leipski} et~al.}{2006}]{leipski2006a}
{Leipski} C.,  {Falcke} H.,  {Bennert} N.,   {H{\"u}ttemeister} S.,  2006,
  Astronomy and Astrophysics, 455, 161

\bibitem[\protect\citeauthoryear{{Lequeux}}{{Lequeux}}{2005}]{lequeux2005a}
{Lequeux} J.,  2005, The Interstellar Medium.
Springer Berlin, Heidelberg

\bibitem[\protect\citeauthoryear{{Liu}, {Zakamska}, {Greene}, {Nesvadba}  \&
  {Liu}}{{Liu} et~al.}{2013}]{liu2013a}
{Liu} G.,  {Zakamska} N.~L.,  {Greene} J.~E.,  {Nesvadba} N. P.~H.,   {Liu} X.,
   2013, The Astrophysical Journal, 436, 2327

\bibitem[\protect\citeauthoryear{{Mel{\'e}ndez}, {Kraemer}, {Weaver}  \&
  {Mushotzky}}{{Mel{\'e}ndez} et~al.}{2011}]{melendez2011a}
{Mel{\'e}ndez} M.,  {Kraemer} S.~B.,  {Weaver} K.~A.,   {Mushotzky} R.~F.,
  2011, The Astrophysical Journal, 738, 6

\bibitem[\protect\citeauthoryear{Merritt \& Ferrarese}{Merritt \&
  Ferrarese}{2001}]{merritt2001a}
Merritt D.,  Ferrarese L.,  2001, \mn@doi [The Astrophysical Journal]
  {10.1086/318372}, 547, 140

\bibitem[\protect\citeauthoryear{Molina et~al.,}{Molina
  et~al.}{2022}]{molina2022a}
Molina J.,  et~al., 2022, \mn@doi [The Astrophysical Journal]
  {10.3847/1538-4357/ac7d4d}, 935, 72

\bibitem[\protect\citeauthoryear{{Mulchaey}, {Wilson}  \&
  {Tsvetanov}}{{Mulchaey} et~al.}{1996}]{mulchaey1996a}
{Mulchaey} J.~S.,  {Wilson} A.~S.,   {Tsvetanov} Z.,  1996, The Astrophysical
  Journal, 467, 197

\bibitem[\protect\citeauthoryear{{Mullaney}, {Alexander}, {Fine}, {Goulding},
  {Harrison}  \& C.}{{Mullaney} et~al.}{2013}]{mullaney2013a}
{Mullaney} J.~R.,  {Alexander} D.~M.,  {Fine} S.,  {Goulding} A.~D.,
  {Harrison} C.~M.,   C. H.~R.,  2013, Monthly Notices of the Royal
  Astronomical Society, 433, 622

\bibitem[\protect\citeauthoryear{{Netzer}}{{Netzer}}{2019}]{netzer2019a}
{Netzer} H.,  2019, Monthly Notices of the Royal Astronomical Society, 488,
  5185

\bibitem[\protect\citeauthoryear{Netzer, Mainieri, Rosati  \&
  Trakhtenbrot}{Netzer et~al.}{2006}]{netzer2006a}
Netzer H.,  Mainieri V.,  Rosati P.,   Trakhtenbrot B.,  2006, A\&A, 453, 525

\bibitem[\protect\citeauthoryear{{Peterson}}{{Peterson}}{1997}]{peterson1997a}
{Peterson} B.~M.,  1997, An introduction to active galactic nuclei, xvi edn.
Cambridge, New York Cambridge University Press

\bibitem[\protect\citeauthoryear{{Rafter}, {Crenshaw}  \& {Wiita}}{{Rafter}
  et~al.}{2009}]{rafter2009a}
{Rafter} S.~E.,  {Crenshaw} D.~M.,   {Wiita} P.~J.,  2009, The Astronomical
  Journal, 137, 42

\bibitem[\protect\citeauthoryear{{Revalski} et~al.,}{{Revalski}
  et~al.}{2018}]{revalski2018a}
{Revalski} M.,  et~al., 2018, The Astrophysical Journal, 867, 88

\bibitem[\protect\citeauthoryear{{Revalski} et~al.,}{{Revalski}
  et~al.}{2021}]{revalski2021a}
{Revalski} M.,  et~al., 2021, \mn@doi [The Astrophysical Journal]
  {https://arxiv.org/abs/2101.06270}, 910, 139

\bibitem[\protect\citeauthoryear{{Revalski} et~al.,}{{Revalski}
  et~al.}{2022}]{revalski2022a}
{Revalski} M.,  et~al., 2022, The Astrophysical Journal, 930, 14

\bibitem[\protect\citeauthoryear{{Schmidt} \& {Green}}{{Schmidt} \&
  {Green}}{1983}]{schmidt1983a}
{Schmidt} M.,  {Green} R.~F.,  1983, The Astrophysical Journal, 269, 352

\bibitem[\protect\citeauthoryear{{Schmitt}, {Donley}, {Antonucci}, {Hutchings}
  \& {Kinney}}{{Schmitt} et~al.}{2003a}]{schmitt2003a}
{Schmitt} H.~R.,  {Donley} J.~L.,  {Antonucci} R. R.~J.,  {Hutchings} J.~B.,
  {Kinney} A.~L.,  2003a, The Astrophysical Journal Supplement Series, 148, 327

\bibitem[\protect\citeauthoryear{{Schmitt}, {Donley}, {Antonucci}, {Hutchings},
  {Kinney}  \& {Pringle}}{{Schmitt} et~al.}{2003b}]{schmitt2003b}
{Schmitt} H.~R.,  {Donley} J.~L.,  {Antonucci} R. R.~J.,  {Hutchings} J.~B.,
  {Kinney} A.~L.,   {Pringle} J.~E.,  2003b, The Astrophysical Journal, 597,
  768

\bibitem[\protect\citeauthoryear{{Seaton}}{{Seaton}}{1979}]{seaton1979a}
{Seaton} M.~J.,  1979, Monthly Notices of the Royal Astronomical Society, 187,
  73

\bibitem[\protect\citeauthoryear{Stern \& Laor}{Stern \&
  Laor}{2012}]{stern2012a}
Stern J.,  Laor A.,  2012, MNRAS, 426, 270

\bibitem[\protect\citeauthoryear{{Storchi-Bergmann} \&
  {Schnorr-Muller}}{{Storchi-Bergmann} \&
  {Schnorr-Muller}}{2019}]{storchi2019a}
{Storchi-Bergmann} T.,  {Schnorr-Muller} A.,  2019, Nature Astronomy, 3, 48

\bibitem[\protect\citeauthoryear{{Storchi-Bergmann}, {Lopes}, {McGregor},
  {Riffel}, {Beck}  \& {Martin}}{{Storchi-Bergmann}
  et~al.}{2010}]{storchi2010a}
{Storchi-Bergmann} T.,  {Lopes} R. D.~S.,  {McGregor} P.~J.,  {Riffel} R.~A.,
  {Beck} T.,   {Martin} P.,  2010, Monthly Notices of the Royal Astronomical
  Society, 402, 819

\bibitem[\protect\citeauthoryear{{Storchi-Bergmann} et~al.,}{{Storchi-Bergmann}
  et~al.}{2018}]{storchi2018a}
{Storchi-Bergmann} T.,  et~al., 2018, The Astrophysical Journal, 868, 14

\bibitem[\protect\citeauthoryear{{Tombesi}, {Cappi}, {Reeves}, {Palumbo},
  {Yaqoob}, {Braito}  \& {Dadina}}{{Tombesi} et~al.}{2010}]{tombesi2010a}
{Tombesi} F.,  {Cappi} M.,  {Reeves} J.~N.,  {Palumbo} G. G.~C.,  {Yaqoob} T.,
  {Braito} V.,   {Dadina} M.,  2010, Astronomy and Astrophysics, 521, A57

\bibitem[\protect\citeauthoryear{{Trindade Falc{\~a}o} et~al.,}{{Trindade
  Falc{\~a}o} et~al.}{2021}]{trindadefalcao2021a}
{Trindade Falc{\~a}o} A.,  et~al., 2021, Monthly Notices of the Royal
  Astronomical Society, 500, 1491

\bibitem[\protect\citeauthoryear{{Woo}, {Bae}, {Son}  \& {Karouzos}}{{Woo}
  et~al.}{2016}]{woo2016a}
{Woo} J.-H.,  {Bae} H.-J.,  {Son} D.,   {Karouzos} M.,  2016, The Astrophysical
  Journal, 817, 108

\makeatother
\end{thebibliography}








\bsp	
\label{lastpage}
\end{document}